\documentclass[twocolumn,superscriptaddress,showpacs,nofootinbib,preprintnumbers,secnumarabic,amssymb, nobibnotes, aps, prd]{revtex4-2}
\usepackage[utf8]{inputenc}
\usepackage{graphicx}
\usepackage{latexsym,amsmath,amssymb,amsthm,lmodern,float,url}
\usepackage{natbib}
\usepackage{color}
\usepackage[usenames,dvipsnames]{xcolor}
\usepackage{microtype}
\usepackage{import}
\usepackage{bbold}
\usepackage[plain]{fancyref}
\usepackage{varioref}
\usepackage{slashed}
\usepackage{multirow}
\usepackage{tikz}
\usepackage{scrextend}
\usepackage{braket}
\usetikzlibrary{shapes}
\usetikzlibrary{positioning}
\usepackage[normalem]{ulem}
\usepackage{todonotes}	
\usepackage{multirow}

\usepackage{qcircuit}

\newcommand{\tab}[1]{Tab.~\ref{tab:#1}}

\usepackage[colorlinks=true,backref=false, linktocpage=true,
citecolor=blue,urlcolor=blue,linkcolor=blue,pdfpagemode=UseOutlines]{hyperref}
\hypersetup{%
  bookmarksnumbered=true,
  pdftitle = {},
  pdfsubject = {},
  pdfauthor = {},
  pdfkeywords = {}
}

\let\Re\undefined

\DeclareMathOperator{\Re}{Re}

\newcommand{\bi}{\mathbb{BI}}
\newcommand{\bo}{\mathbb{BO}}
\newcommand{\btt}{\mathbb{BT}}
\DeclareMathOperator{\saa}{\Sigma(108)}
\DeclareMathOperator{\sab}{\Sigma(216)}
\DeclareMathOperator{\sac}{\Sigma(648)}
\DeclareMathOperator{\sad}{\Sigma(1080)}

\newcommand{\one}{\mathbf{1}}

\newcommand{\three}{\mathbf{3}}

\newcommand{\five}{\mathbf{5}}
\newcommand{\six}{\mathbf{6}}
\newcommand{\eight}{\mathbf{8}}
\newcommand{\nine}{\mathbf{9}}
\newcommand{\ten}{\mathbf{10}}
\newcommand{\ften}{\mathbf{15}}

\definecolor{blugrn}{RGB}{0,158,115}

\begin{document}
\preprint{FERMILAB-PUB-24-0177-T}

\title{
Digitization and subduction of $SU(N)$ gauge theories
}
\author{Beno\^{i}t Assi }
\email{bassi@fnal.gov}
\affiliation{Fermi National Accelerator Laboratory, Batavia, Illinois, 60510, USA}
\author{Henry Lamm}
\email{hlamm@fnal.gov}
\affiliation{Fermi National Accelerator Laboratory, Batavia, Illinois, 60510, USA}

\date{\today}

\begin{abstract}
The simulation of lattice gauge theories on quantum computers necessitates digitizing gauge fields. One approach involves substituting the continuous gauge group with a discrete subgroup, but the implications of this approximation still need to be clarified. To gain insights, we investigate the subduction of \( SU(2) \) and \( SU(3) \) to discrete crystal-like subgroups. Using classical lattice calculations, we show that subduction offers valuable information based on subduced direct sums, helping us identify additional terms to incorporate into the lattice action that can mitigate the effects of digitization. Furthermore, we compute the static potentials of all irreducible representations of \( \Sigma(360 \times 3) \) at a fixed lattice spacing. Our results reveal a percent-level agreement with the Casimir scaling of \( SU(3) \) for irreducible representations that subduce to a single \( \Sigma(360 \times 3) \) irreducible representation. This provides a diagnostic measure of approximation quality, as some irreducible representations closely match the expected results while others exhibit significant deviations.
\end{abstract}

\maketitle

\section{Introduction}

Digital quantum computers offer promising opportunities for exploring lattice gauge theories (LGTs) in regimes that classical computers struggle with due to sign problems, such as real-time dynamics or the properties of matter at finite density \cite{Aarts_2016, Philipsen:2007aa, forcr2010simulating, 10.1143/PTP.110.615, Gattringer:2016kco, Alexandru:2018ddf, Alexandru:2020wrj, Troyer:2004ge}. However, these devices also bring significant challenges, primarily the limited availability of qubits and the shallow circuit depth before noise overtakes the calculations. This constraint is reminiscent of the early days of Euclidean LGT on classical computers, where storing floating-point representations of \( SU(3) \) was computationally prohibitive. This led to research on approximations \cite{Creutz:1979zg, Kogut:1979wt, Lisboa:1982ji, Lisboa:1982jj} and gauge-fixing \cite{Giusti:2001xf}.

Many digitizations have been proposed to reduce this high cost~\cite{Zohar:2012ay,Zohar:2012xf,Zohar:2013zla,Zohar:2014qma,Zohar:2015hwa,Zohar:2016iic,Klco:2019evd,Ciavarella:2021nmj,Bender:2018rdp,Liu:2020eoa,Hackett:2018cel,Alexandru:2019nsa,Yamamoto:2020eqi,Haase:2020kaj,Armon:2021uqr,PhysRevD.99.114507,Bazavov:2015kka,Zhang:2018ufj,Unmuth-Yockey:2018ugm,Unmuth-Yockey:2018xak,Kreshchuk:2020dla,Kreshchuk:2020aiq,Raychowdhury:2018osk,Raychowdhury:2019iki,Davoudi:2020yln,Wiese:2014rla,Luo:2019vmi,Brower:2020huh,Mathis:2020fuo,Singh:2019jog,Singh:2019uwd,Buser:2020uzs,Bhattacharya:2020gpm,Barata:2020jtq,Kreshchuk:2020kcz,Ji:2020kjk,Bauer:2021gek,Gustafson:2021qbt,Halimeh:2021vzf,Hartung:2022hoz,Osborne:2022jxq,Grabowska:2022uos,Murairi:2022zdg,Bauer:2023jvw,Gustafson:2023kvd,Alexandru:2023qzd,Halimeh:2023wrx,Fromm:2023bit,Fromm:2023npm,Ciavarella:2024fzw,Bergner:2024qjl}. Each makes choices that break symmetries~\cite{Zache:2023dko,Hackett:2018cel} complicating the extrapolation back to original theory~\cite{Hasenfratz:2001iz,Caracciolo:2001jd,Hasenfratz:2000hd,PhysRevE.57.111,PhysRevE.94.022134,car_article,Singh:2019jog,Singh:2019uwd,Bhattacharya:2020gpm,Zhou:2021qpm,Caspar:2022llo,Jakobs:2023lpp,Hartung:2022hoz}. Moreover, the effectiveness of the digitization can be spacetime dimension-dependent~\cite{Davoudi:2020yln,Zohar:2021nyc,Alam:2021uuq,Gustafson:2022xdt,Davoudi:2022xmb}. Further, some digitizations can be simulated on classical computers without sign problems, allowing for nonperturbative Euclidean simulations or stochastic state preparation~\cite{Lamm:2018siq,Harmalkar:2020mpd,Gustafson:2020yfe,Saroni:2023uob}. Currently, there is a significant gap in our understanding regarding the resource demands, potential errors, and the feasibility of achieving continuum limits with these techniques.

In this paper, we will consider the discrete subgroup approximation, which uses comparatively few qubits (or bits) and shallow circuit depth by avoiding fixed-point arithmetic. This digitization was first studied in Euclidean simulations for reducing classical resources~\cite{Creutz:1979zg,Creutz:1982dn,Bhanot:1981xp,Petcher:1980cq,Bhanot:1981pj} including with dynamical fermions~\cite{Weingarten:1980hx,Weingarten:1981jy}. Today, this approximation's potential for use in quantum computing is under investigation~\cite{Bender:2018rdp,Hackett:2018cel,Alexandru:2019nsa,Yamamoto:2020eqi,Ji:2020kjk,Haase:2020kaj,Carena:2021ltu,Armon:2021uqr,Gonzalez-Cuadra:2022hxt,Gustafson:2022xdt,Alam:2021uuq,Fromm:2022vaj,Carena:2022hpz,Charles:2023zbl,Gustafson:2023kvd,Gustafson:2023swx,Ballini:2023ljs,Carena:2024dzu,Gustafson:2024kym}. 

This work will focus on how the subduction of the continuous $SU(2)$ and $SU(3)$ groups to their discrete subgroups plays a role in the digitization error. We will use the notation of $\mathcal{G}$ to represent a continuous group and $\mathbb{G}$ to denote any of its discrete subgroups. Subduction provides a mapping from irreducible representations (IR) of one group to one or a direct sum of IRs of a subgroup. This builds on previous work, which investigated how reducing continuous spacetime to a cubic lattice corresponds to the subduction of $SU(2)$ to the binary octahedral group $\bo$ -- breaking in the rotational symmetry of LGT~\cite{Johnson:1982yq,Moore:2005dw}.  From this understanding, it has been possible to construct an improved operator with reduced signal-to-noise and excited state contamination~\cite{Basak:2007kj,Edwards:2011jj,Egerer:2021ymv,Detmold:2024ifm}. We extend these ideas to gauge digitization by providing subduction tables for crystal-like subgroups of both $SU(2)$ and $SU(3)$ and use this to gain insight by analyzing two lattice observables -- the lattice energy density and Casimir scaling of the nonperturbative static potentials. This quantitative understanding of the breaking of the continuous gauge symmetry represents a starting point to systematically interpret the discrete subgroups as continuous groups broken by a Higgs mechanism~\cite{Horn:1979fy,Kogut:1980qb,Fradkin:1978dv,Ovrut:1977cn,Merle:2011vy,Luhn:2011ip,Vileta:2022jou}.

This paper is organized as follows. Sec.~\ref{sec:su2} and ~\ref{sec:su3} summarizing key properties of, respectively, the $SU(2)$ and $SU(3)$ and their crystal-like subgroups. In Sec.~\ref{sec:subd}, we determine the subductions and inductions between the continuous and discrete groups. Following that, in Sec.~\ref{sec:num}, numerical results are presented for the lattice energy density and Casimir scaling of the static potentials; we then conclude in Sec.~\ref{sec:con}.

\section{\texorpdfstring{$SU(2)$}{SU(2)}}\label{sec:su2}

The infinite set of IRs of $SU(2)$ are indexed by a half-integer $j$, and their dimensionality is given by $d=2j+1$. In a given IR, the character of group elements further depends on a rotation angle $\theta$:
\begin{equation}
\label{eq:su2char}
    \chi^{(j)}(\theta)=\frac{\sin\left(\left[j+\frac{1}{2}\right]\theta\right)}{\sin\left(\frac{\theta}{2}\right)}.
\end{equation}

\begin{table}
    \caption{Character Table of $\mathbb{BT}$\footnote{A typo in the 7th line of \cite{Gustafson:2022xdt} has been corrected here.}}
    \label{tab:charbt}
    \begin{tabular}{c||cc|c|cc|cc}
Size & 1 & 1 & 6 & 4 & 4 & 4 & 4 \\
Order & 1 & 2 & 4 & 6 & 6 & 3 & 3 \\
\hline\hline               
$A_0$ & 1 & 1 & 1 & 1 & 1 & 1 & 1 \\
$E_1$ & 1 & 1 & 1 & $\omega$ & $\omega^2$ & $\omega^2$ & $\omega$ \\
$E_2$ & 1 & 1 & 1 & $\omega^2$ & $\omega$ & $\omega$ & $\omega^2$ \\ \hline
$H$ & 2 & $-2$ & 0 & 1 & 1 & $-1$ & $-1$ \\
$G_1$ & 2 & $-2$ & 0 & $\omega$ & $\omega^2$ & $-\omega^2$ & $-\omega$ \\
$G_2$ & 2 & $-2$ & 0 & $\omega^2$ & $\omega$ & $-\omega$ & $-\omega^2$ \\ \hline
$T$ & 3 & 3 & $-1$ & 0 & 0 & 0 & 0 \\       
    \end{tabular}
\end{table}

$SU(2)$ has three crystal-like subgroups: the 24-element Binary Tetrahedral group ($\btt$), the 48-element Binary Octahedral group ($\bo$), and the 120-element Binary Icosahedral group ($\mathbb{BI}$).  The character tables for each group are given by Tables~\ref{tab:charbt},~\ref{tab:charbo}, and~\ref{tab:charbi} respectively and will be used below to derive the subduction of $SU(2)$. 

\begin{table}
    \caption{Character table of $\mathbb{BO}$}
    \label{tab:charbo}
    \begin{tabular}{c||cc|c|c|cc|cc}
Size & 1 & 1 & 12 & 6 & 8 & 8 & 6 & 6\\
Ord. & 1 & 2 & 4 & 4 & 6 & 3 & 8 & 8 \\
\hline\hline               
$A_1$ & 1 & 1 & 1 & 1 & 1 & 1 & 1 & 1\\
$A_2$ & 1 & 1 & $-1$ & 1 & 1 & 1 & $-1$ & $-1$\\ \hline
$E$ & 2 & $2$ & 0 &2 & $-1$ & $-1$ & 0 & 0\\
$G_1$ & 2 & $-2$ & 0 & 0& 1 & $-1$ & $-\sqrt{2}$ & $\sqrt{2}$ \\
$G_2$ & 2 & $-2$ & 0 & 0& 1 & $-1$ & $\sqrt{2}$ & $-\sqrt{2}$ \\ \hline
$T_1$ & 3 & 3 & $-1$ & $-1$ & 0 & 0 & 1 & 1\\
$T_2$ & 3 & 3 & 1 & $-1$ & 0 & 0 & $-1$ & $-1$\\ \hline
$H$ & 4 & $-4$ & 0 & 0 & $-1$ & 1 & 0 & 0\\
    \end{tabular}
\end{table}

\begin{table}
    \caption{Character table of $\mathbb{BI}$}
    \label{tab:charbi}
    \begin{tabular}{c||cc|c|c|cc|c|cc}
Size & 1 & 1 & 20 & 30 & 12 & 12 & 20 & 12 & 12\\
Ord. & 1 & 2 & 3 & 4 & 5 & 5 & 6 & 10 & 10\\
\hline\hline               
$A_0$    &1& 1& 1& 1& 1& 1& 1& 1& 1\\ \hline
$E_1$    & 2& -2 & -1& 0& $\frac{-1+\sqrt{5}}{2}$& $\frac{-1-\sqrt{5}}{2}$& 1& $\frac{1+\sqrt{5}}{2}$& $\frac{1-\sqrt{5}}{2}$\\
$E_2$    & 2& -2 & -1& 0& $\frac{-1-\sqrt{5}}{2}$& $\frac{-1+\sqrt{5}}{2}$& 1& $\frac{1-\sqrt{5}}{2}$& $\frac{1+\sqrt{5}}{2}$\\ \hline
$T_1$    &3& 3& 0 & -1& $\frac{1-\sqrt{5}}{2}$& $\frac{1+\sqrt{5}}{2}$& 0& $\frac{1+\sqrt{5}}{2}$& $\frac{1-\sqrt{5}}{2}$\\
$T_2$    & 3& 3& 0& -1& $\frac{1+\sqrt{5}}{2}$& $\frac{1-\sqrt{5}}{2}$& 0& $\frac{1-\sqrt{5}}{2}$& $\frac{1+\sqrt{5}}{2}$\\ \hline
$G_1$    & 4 & 4 & 1& 0& -1& -1& 1& -1& -1 \\
$G_2$    & 4& -4& 1& 0& -1& -1& -1& 1& 1\\ \hline
$H$    &5& 5& -1& 1& 0& 0& -1& 0& 0\\ \hline
$J$    & 6& -6& 0& 0& 1& 1& 0& -1& -1\\
    \end{tabular}
\end{table}

\section{\texorpdfstring{$SU(3)$}{SU(3)}}\label{sec:su3}
The infinite set of IRs of $SU(3)$, $ \chi^{(p,q)} $, are indexed by non-negative integers $p,q$, which correspond to the highest weights of the representation, reflecting the symmetry properties of particle states under $SU(3)$ transformations. The dimensionality of $ \chi^{(p,q)} $ is $d=\frac{1}{2}(p+1)(q+1)(p+q+2)$. In a given IR, the character of group elements further depends on rotation angles $\theta,\phi$:
\begin{equation}
\chi^{(p,q)}=e^{i(p+2q)\theta}\sum_{k=q}^{p+q}\sum_{l=0}^q e^{-3i(k+l)\frac{\theta}{2}}\frac{\sin\left(\left[k-l+1\right]\frac{\phi}{2}\right)}{\sin\left(\frac{\phi}{2}\right)},
\end{equation}
We classify the discrete subgroups of $SU(3)$ we are interested in and their associated IRs. The finite, non-Abelian crystal-like subgroups of $SU(3)$ with a $\mathbb{Z}_3$ center are $\Sigma(108)$, $\Sigma(216)$, $\Sigma(648)$, and $\Sigma(1080)$, where the number in parentheses indicates the number of elements. We provide the character tables for these discrete subgroups in Tables~\ref{tab:cts180},~\ref{tab:cts216},~\ref{tab:cts648} and~\ref{tab: s1080-character-evaluation}, respectively.

\begin{table}
\caption{Character table of $\saa$ with $\omega=e^{2\pi i/3}$.
\label{tab:cts180}}
\begin{tabular}{c||ccc|c|c|ccc|ccc|ccc}
Size & 1 & 1 & 1 & 12 & 12 & 9 & 9 & 9 &
9 & 9 & 9 & 9 & 9 & 9 \\
Ord. & 1 & 3 & 3 & 3 & 3 & 2 & 6 & 6 & 4 & 12 & 12 & 4
& 12 & 12 \\ 
\hline \hline
$\one^{(0)}$ & $1$ & $1$ & $1$ & $1$ & $1$ 
& $1$ & $1$ & $1$         
& $1$ & $1$ & $1$         
& $1$ & $1$ & $1$ \\      
$\one^{(1)}$ & $1$ & $1$ & $1$ & $1$ & $1$ 
& $\text{-}1$ & $\text{-}1$ & $\text{-}1$
& $i$ & $i$ & $i$ 
& $\text{-}i$ & $\text{-}i$ & $\text{-}i$ \\
$\one^{(2)}$ & $1$ & $1$ & $1$ & $1$ & $1$ 
& $1$ & $1$ & $1$
& $\text{-}1$ & $\text{-}1$ & $\text{-}1$ 
& $\text{-}1$ & $\text{-}1$ & $\text{-}1$ \\
$\one^{(3)}$ & $1$ & $1$ & $1$ & $1$ & $1$ 
& $\text{-}1$ & $\text{-}1$ & $\text{-}1$
& $\text{-}i$ & $\text{-}i$ & $\text{-}i$ & 
$i$ & $i$ & $i$ \\
\hline
$\three^{(0)}$ & $3$ & $3 \omega$ & $3 \omega^2$ &
$0$ & $0$ &
$\text{-}1$ & $\text{-}\omega$ & $\text{-}\omega^2$ &
$1$ & $\omega$ & $\omega^2$ &
$1$ & $\omega$ & $\omega^2$ \\
$\three^{(1)}$ & $3$ & $3 \omega$ & $3 \omega^2$ &
$0$ & $0$ &
$1$ & $\omega$ & $\omega^2$ &
$i$ & $i \omega$ & $i \omega^2$ &
$\text{-}i$ & $\text{-}i \omega$ & $\text{-}i \omega^2$ \\
$\three^{(2)}$ & $3$ & $3 \omega$ & $3 \omega^2$ &
$0$ & $0$ &
$\text{-}1$ & $\text{-}\omega$ & $\text{-}\omega^2$ &
$\text{-}1$ & $\text{-}\omega$ & $\text{-}\omega^2$ &
$\text{-}1$ & $\text{-}\omega$ & $\text{-}\omega^2$ \\
$\three^{(3)}$ & $3$ & $3 \omega$ & $3 \omega^2$ &
$0$ & $0$ &
$1$ & $\omega$ & $\omega^2$ &
$\text{-}i$ & $\text{-}i \omega$ & $\text{-}i \omega^2$ &
$i$ & $i \omega$ & $i \omega^2$ \\
$\three^{(0)*}$  & $3$ & $3 \omega^2$ & $3 \omega$ &
$0$ & $0$ &
$\text{-}1$ & $\text{-}\omega^2$ & $\text{-}\omega$ &
$1$ & $\omega^2$ & $\omega$ &
$1$ & $\omega^2$ & $\omega$ \\
$\three^{(1)^*}$ & $3$ & $3 \omega^2$ & $3 \omega$ &
$0$ & $0$ &
$1$ & $\omega^2$ & $\omega$ &
$\text{-}i$ & $\text{-}i \omega^2$ & $\text{-}i \omega$ &
$i$ & $i \omega^2$ & $i \omega$ \\
$\three^{(2)*}$  & $3$ & $3 \omega^2$ & $3 \omega$ &
$0$ & $0$ &
$\text{-}1$ & $\text{-}\omega^2$ & $\text{-}\omega$ &
$\text{-}1$ & $\text{-}\omega^2$ & $\text{-}\omega$ &
$\text{-}1$ & $\text{-}\omega^2$ & $\text{-}\omega$ \\
$\three^{(3)*}$ & $3$ & $3 \omega^2$ & $3 \omega$ &
$0$ & $0$ &
$1$ & $\omega^2$ & $\omega$ &
$i$ & $i \omega^2$ & $i \omega$ &
$\text{-}i$ & $\text{-}i \omega^2$ & $\text{-}i \omega$ \\
\hline
$\mathbf{4}$ & $4$ & $4$ & $4$ &
$1$ & $\text{-}2$ &
$0$ & $0$ & $0$ &
$0$ & $0$ & $0$ &
$0$ & $0$ & $0$ \\
$\mathbf{4}^\prime$ & $4$ & $4$ & $4$ &
$\text{-}2$ & $1$ &
$0$ & $0$ & $0$ &
$0$ & $0$ & $0$ &
$0$ & $0$ & $0$ \\
\end{tabular}
\end{table}

\begin{table*}
\caption{Character table of $\sab$ with $\omega=e^{2\pi i/3}$.
\label{tab:cts216}}
\begin{tabular}{c||ccc|c|ccc|ccc|ccc|ccc}
Size & 1&1&1 & 24 & 9&9&9 & 18 & 18 &18 &18 &18 &18 &18 &18 &18\\
Ord & 1 & 3 & 3 & 3 & 2 & 6 & 6 & 4 & 12 & 12 & 4 & 12 & 12 &
12 & 12 & 4 \\
\hline \hline
$\mathbf{1}^{(0)}$ & $1$ & $1$ & $1$ & $1$ & $1$ 
& $1$ & $1$ & $1$
& $1$ & $1$ & $1$
& $1$ & $1$ & $1$ & $1$ & $1$ \\
$\mathbf{1}^{(1)}$ & $1$ & $1$ & $1$ & $1$ & $1$ 
& $1$ & $1$ & $\text{-}1$
& $\text{-}1$ & $\text{-}1$ & $1$ 
& $1$ & $1$ & $\text{-}1$ & $\text{-}1$ & $\text{-}1$ \\
$\mathbf{1}^{(2)}$ &
$1$ & $1$ & $1$ & 
$1$ &
$1$ & $1$ & $1$ &
$1$ & $1$ & $1$ &
$\text{-}1$ & $\text{-}1$ & $\text{-}1$ &
$\text{-}1$ & $\text{-}1$ & $\text{-}1$ \\
$\mathbf{1}^{(3)}$ &
$1$ & $1$ & $1$ & 
$1$ &
$1$ & $1$ & $1$ &
$\text{-}1$ & $\text{-}1$ & $\text{-}1$ &
$\text{-}1$ & $\text{-}1$ & $\text{-}1$ &
$1$ & $1$ & $1$ \\
\hline
$\mathbf{2}$ &
$2$ & $2$ & $2$ & 
$2$ &
$\text{-}2$ & $\text{-}2$ & $\text{-}2$ &
$0$ & $0$ & $0$ &
$0$ & $0$ & $0$ &
$0$ & $0$ & $0$ \\
\hline
$\mathbf{3}^{(0)}$ &
$3$ & $3\omega$ & $3\omega^2$ & 
$0$ &
$\text{-}1$ & $\text{-}\omega$ & $\text{-}\omega^2$ &
$1$ & $\omega$ & $\omega^2$ &
$1$ & $\omega$ & $\omega^2$ &
$\omega$ & $\omega^2$ & $1$ \\
$\mathbf{3}^{(1)}$ &
$3$ & $3\omega$ & $3\omega^2$ & 
$0$ &
$\text{-}1$ & $\text{-}\omega$ & $\text{-}\omega^2$ &
$\text{-}1$ & $\text{-}\omega$ & $\text{-}\omega^2$ &
$1$ & $\omega$ & $\omega^2$ &
$\text{-}\omega$ & $\text{-}\omega^2$ & $\text{-}1$ \\
$\mathbf{3}^{(2)}$ &
$3$ & $3\omega$ & $3\omega^2$ & 
$0$ &
$\text{-}1$ & $\text{-}\omega$ & $\text{-}\omega^2$ &
$1$ & $\omega$ & $\omega^2$ &
$\text{-}1$ & $\text{-}\omega$ & $\text{-}\omega^2$ &
$\text{-}\omega$ & $\text{-}\omega^2$ & $\text{-}1$ \\
$\mathbf{3}^{(3)}$ &
$3$ & $3\omega$ & $3\omega^2$ & 
$0$ &
$\text{-}1$ & $\text{-}\omega$ & $\text{-}\omega^2$ &
$\text{-}1$ & $\text{-}\omega$ & $\text{-}\omega^2$ &
$\text{-}1$ & $\text{-}\omega$ & $\text{-}\omega^2$ &
$\omega$ & $\omega^2$ & $1$ \\
$\mathbf{3}^{(0)*}$ &
$3$ & $3\omega^2$ & $3\omega$ & 
$0$ &
$\text{-}1$ & $\text{-}\omega^2$ & $\text{-}\omega$ &
$1$ & $\omega^2$ & $\omega$ &
$1$ & $\omega^2$ & $\omega$ &
$\omega^2$ & $\omega$ & $1$ \\
$\mathbf{3}^{(1)*}$ &
$3$ & $3\omega^2$ & $3\omega$ & 
$0$ &
$\text{-}1$ & $\text{-}\omega^2$ & $\text{-}\omega$ &
$\text{-}1$ & $\text{-}\omega^2$ & $\text{-}\omega$ &
$1$ & $\omega^2$ & $\omega$ &
$\text{-}\omega^2$ & $\text{-}\omega$ & $\text{-}1$ \\
$\mathbf{3}^{(2)*}$ &
$3$ & $3\omega^2$ & $3\omega$ & 
$0$ &
$\text{-}1$ & $\text{-}\omega^2$ & $\text{-}\omega$ &
$1$ & $\omega^2$ & $\omega$ &
$\text{-}1$ & $\text{-}\omega^2$ & $\text{-}\omega$ &
$\text{-}\omega^2$ & $\text{-}\omega$ & $\text{-}1$ \\
$\mathbf{3}^{(3)*}$ &
$3$ & $3\omega^2$ & $3\omega$ & 
$0$ &
$\text{-}1$ & $\text{-}\omega^2$ & $\text{-}\omega$ &
$\text{-}1$ & $\text{-}\omega^2$ & $\text{-}\omega$ &
$\text{-}1$ & $\text{-}\omega^2$ & $\text{-}\omega$ &
$\omega^2$ & $\omega$ & $1$ \\
\hline
$\mathbf{6}$ &
$6$ & $6\omega$ & $6\omega^2$ & 
$0$ &
$2$ & $2\omega$ & $2\omega^2$ &
$0$ & $0$ & $0$ &
$0$ & $0$ & $0$ &
$0$ & $0$ & $0$ \\
$\mathbf{6^{\ast}}$ &
$6$ & $6\omega^2$ & $6\omega$ & 
$0$ &
$2$ & $2\omega^2$ & $2\omega$ &
$0$ & $0$ & $0$ &
$0$ & $0$ & $0$ &
$0$ & $0$ & $0$ \\
\hline
$\mathbf{8}$ &
$8$ & $8$ & $8$ & 
$\text{-}1$ &
$0$ & $0$ & $0$ &
$0$ & $0$ & $0$ &
$0$ & $0$ & $0$ &
$0$ & $0$ & $0$ \\
\end{tabular}
\end{table*}

\begin{table*}
\centering
\caption{Character table of $\sac$ with $\omega=e^{2\pi i/3}$,
  $\rho \equiv e^{2\pi i/9}$, $\sigma \equiv \rho(1+2\omega)$
\label{tab:cts648}}
\begin{tabular}{c||ccc|c|ccc|ccc|c|c|ccc|ccc|ccc|ccc}
Size & 1 & 1 & 1 & 24 & 9 & 9 & 9 & 54 &
54 & 54 & 72 & 72 & 12 & 12 & 12 & 12 & 12 & 12 & 36 & 36 &
36 & 36 & 36 & 36 \\
Ord. & 1 & 3 & 3 & 3 & 2 & 6 & 6 & 4 & 12 & 12 & 3 & 3 & 9 & 9 & 9 & 9 & 9 & 9 & 18 & 18 & 18 & 18 & 18 & 18 \\
\hline \hline
$\mathbf{1}^{(0)}$ & 1 & 1 & 1 & 1 & 1 & 1 & 1 & 1 & 1 & 1 & 1 & 1& 1 & 1 & 1 & 1 & 1 & 1 & 1 & 1 & 1 & 1 & 1 & 1 \\
$\mathbf{1}^{(1)}$ & 1 & 1 & 1 & 1 & 1 & 1 & 1 & 1 & 1 & 1 & 
$\omega$ & $\omega^2$ &  $\omega$ & $\omega$ & $\omega$ &
$\omega^2$ &  $\omega^2$ &  $\omega^2$ & $\omega$ &   $\omega$ &
$\omega$ &  $\omega^2$ &  $\omega^2$ &  $\omega^2$ \\
$\mathbf{1}^{(2)}$ & 1 & 1 & 1 & 1 & 1 & 1 & 1 & 1 & 1 & 1 &
$\omega^2$ & $\omega$ &  $\omega^2$ &  $\omega^2$ &  $\omega^2$ &
$\omega$ &   $\omega$ &   $\omega$ &  $\omega^2$ &  $\omega^2$ &
$\omega^2$ &   $\omega$ &   $\omega$ &   $\omega$ \\ 
\hline
$\mathbf{2}^{(0)}$ & 2 & 2 & 2 & 2 & $\text{-}2$ & $\text{-}2$ & $\text{-}2$ &   0 &    0 & 0 & $\text{-}1$ & $\text{-}1$ & $\text{-}1$ & $\text{-}1$ & $\text{-}1$ & $\text{-}1$ & $\text{-}1$ & $\text{-}1$ & 1 & 1 &
1 & 1 & 1 & 1 \\
$\mathbf{2}^{(1)}$ & 2 & 2 & 2 & 2 & $\text{-}2$ & $\text{-}2$ & $\text{-}2$ &   0 &    0 &0 &    $\text{-}\omega$ &   $\text{-}\omega^2$ & $\text{-}\omega$ &    $\text{-}\omega$ &    $\text{-}\omega$ & $\text{-}\omega^2$ &   $\text{-}\omega^2$ &   $\text{-}\omega^2$ &   $\omega$ &   $\omega$ & $\omega$ &  $\omega^2$ &  $\omega^2$ &  $\omega^2$ \\ 
$\mathbf{2}^{(2)}$ & 2 & 2 & 2 & 2 & $\text{-}2$ & $\text{-}2$ & $\text{-}2$ &   0 &    0 & 0 &   $\text{-}\omega^2$ &    $\text{-}\omega$ & $\text{-}\omega^2$ & $\text{-}\omega^2$ & $\text{-}\omega^2$ & $\text{-}\omega$ &    $\text{-}\omega$ &    $\text{-}\omega$ &  $\omega^2$ &  $\omega^2$ & $\omega^2$ &   $\omega$ &   $\omega$ &   $\omega$ \\ 
\hline
$\mathbf{3}^{(a)}$ & 3 & 3 & 3 & 3 & 3 & 3 & 3 & $\text{-}1$ & $\text{-}1$ & $\text{-}1$ & 0 & 0 & 0 & 0 & 0 & 0 & 0 & 0 & 0 & 0 & 0 & 0 & 0 & 0 \\
$\mathbf{3}^{(0)}$ & 3 &   $3\omega$ &  $3\omega^2$ & 0 & $\text{-}1$ & $\text{-}\omega$ &  $\text{-}\omega^2$ & 1 &   $\omega$ &  $\omega^2$ & 0 &  0 &   $\omega^2\sigma^{\ast}$ &   $\sigma^{\ast}$ & $\omega\sigma^{\ast}$ &  $\sigma$ &    $\omega\sigma$ & $\omega^2\sigma$ &   $\text{-}\rho^2$ &   $\text{-}\rho^{4*}$ & $\text{-}\rho^{\ast}$ &    $\text{-}\rho^{2*}$ &    $\text{-}\rho$ &    $\text{-}\rho^{4}$ \\
$\mathbf{3}^{(1)}$ & 3 &   $3\omega$ &  $3\omega^2$ &   0 & $\text{-}1$ & $\text{-}\omega$ &  $\text{-}\omega^2$ & 1 &   $\omega$ &  $\omega^2$ &    0 &    0 &   $\sigma^{\ast}$ &   $\omega\sigma^{\ast}$ & $\omega^2\sigma^{\ast}$ &    $\omega^2\sigma$ &    $\sigma$ & $\omega\sigma$ &   $\text{-}\rho^{4*}$ &   $\text{-}\rho^{\ast}$ & $\text{-}\rho^2$ &    $\text{-}\rho^{4}$ &    $\text{-}\rho^{2*}$ & $\text{-}\rho$ \\
$\mathbf{3}^{(2)}$ & 3 &   $3\omega$ &  $3\omega^2$ &   0 & $\text{-}1$ & $\text{-}\omega$ &  $\text{-}\omega^2$ & 1 &   $\omega$ &  $\omega^2$ &    0 &    0 &   $\omega\sigma^{\ast}$ & $\omega^2\sigma^{\ast}$ &   $\sigma^{\ast}$ &    $\omega\sigma$ & $\omega^2\sigma$ &    $\sigma$ &   $\text{-}\rho^{\ast}$ & $\text{-}\rho^2$ &   $\text{-}\rho^{4*}$ &    $\text{-}\rho$ & $\text{-}\rho^{4}$ & $\text{-}\rho^{2*}$ \\
$ \mathbf{3}^{(0)*}$ & 3 &  $3\omega^2$ &   $3\omega$ & 0 & $\text{-}1$ &  $\text{-}\omega^2$ &   $\text{-}\omega$ & 1 &  $\omega^2$ &   $\omega$ & 0 & 0 & $\omega\sigma$ & $\sigma$ & $\omega^2\sigma$ &   $\sigma^{\ast}$ &   $\omega^2\sigma^{\ast}$ & $\omega\sigma^{\ast}$ &    $\text{-}\rho^{2*}$ &    $\text{-}\rho^{4}$ & $\text{-}\rho$ &   $\text{-}\rho^2$ &   $\text{-}\rho^{\ast}$ & $\text{-}\rho^{4*}$ \\ 
$ \mathbf{3}^{(1)*}$ & 3 &  $3\omega^2$ &   $3\omega$ & 0 & $\text{-}1$ &  $\text{-}\omega^2$ &   $\text{-}\omega$ & 1 &  $\omega^2$ &   $\omega$ & 0 & 0 & $\sigma$ &    $\omega^2\sigma$ & $\omega\sigma$ &   $\omega\sigma^{\ast}$ &   $\sigma^{\ast}$ & $\omega^2\sigma^{\ast}$ &    $\text{-}\rho^{4}$ &  $\text{-}\rho$ & $\text{-}\rho^{2*}$ &   $\text{-}\rho^{4*}$ &   $\text{-}\rho^2$ & $\text{-}\rho^{\ast}$ \\
$ \mathbf{3}^{(2)*}$ & 3 &  $3\omega^2$ &   $3\omega$ & 0 & $\text{-}1$ &  $\text{-}\omega^2$ &   $\text{-}\omega$ & 1 &  $\omega^2$ &   $\omega$ & 0 & 0 &    $\omega^2\sigma$ & $\omega\sigma$ &    $\sigma$ &   $\omega^2\sigma^{\ast}$ & $\omega\sigma^{\ast}$ &   $\sigma^{\ast}$ &    $\text{-}\rho$ & $\text{-}\rho^{2*}$ &    $\text{-}\rho^{4}$ &   $\text{-}\rho^{\ast}$ & $\text{-}\rho^{4*}$ &   $\text{-}\rho^2$ \\
\hline
$\mathbf{6}^{(0)}$ & 6 & $6\omega$ &   $6\omega^2$ &   0 & 2 & $2\omega$ &  $2\omega^2$ & 0 & 0 & 0 & 0 & 0 &  $\text{-}\omega\sigma^{\ast}$ & $\text{-}\omega^2\sigma^{\ast}$ &  $\text{-}\sigma^{\ast}$ &   $\text{-}\omega\sigma$ & $\text{-}\omega^2\sigma$ &   $\text{-}\sigma$ &   $\text{-}\rho^{\ast}$ & $\text{-}\rho^2$ &   $\text{-}\rho^{4*}$ &    $\text{-}\rho$ & $\text{-}\rho^{4}$ &    $\text{-}\rho^{2*}$ \\
$\mathbf{6}^{(1)}$ & 6 &    $6\omega$ &   $6\omega^2$ &   0 & 2 & $2\omega$ &  $2\omega^2$ &   0 &    0 &    0 &    0 & 0 &  $\text{-}\omega^2\sigma^{\ast}$ &  $\text{-}\sigma^{\ast}$ & $\text{-}\omega\sigma^{\ast}$ &   $\text{-}\sigma$ &   $\text{-}\omega\sigma$ &
$\text{-}\omega^2\sigma$ &   $\text{-}\rho^2$ &   $\text{-}\rho^{4*}$ &   $\text{-}\rho^{\ast}$ &    $\text{-}\rho^{2*}$ &  $\text{-}\rho$ & $\text{-}\rho^{4}$ \\
$\mathbf{6}^{(2)}$ & 6 &    $6\omega$ &   $6\omega^2$ &   0 & 2 & $2\omega$ &  $2\omega^2$ &   0 &    0 &    0 &    0 & 0 &  $\text{-}\sigma^{\ast}$ &  $\text{-}\omega\sigma^{\ast}$ & $\text{-}\omega^2\sigma^{\ast}$ &   $\text{-}\omega^2\sigma$ &   $\text{-}\sigma$ &
$\text{-}\omega\sigma$ &   $\text{-}\rho^{4*}$ &   $\text{-}\rho^{\ast}$ & $\text{-}\rho^2$ &    $\text{-}\rho^{4}$ &    $\text{-}\rho^{2*}$ & $\text{-}\rho$ \\
$ \mathbf{6}^{(0)*}$ & 6 &   $6\omega^2$ &    $6\omega$ & 0 & 2 &  $2\omega^2$ & $2\omega$ & 0 & 0 & 0 & 0 & 0 & $\text{-}\omega^2\sigma$ & $\text{-}\omega\sigma$ &   $\text{-}\sigma$ &  $\text{-}\omega^2\sigma^{\ast}$ & $\text{-}\omega\sigma^{\ast}$ &  $\text{-}\sigma^{\ast}$ &    $\text{-}\rho$ & $\text{-}\rho^{2*}$ &    $\text{-}\rho^{4}$ &   $\text{-}\rho^{\ast}$ & $\text{-}\rho^{4*}$ &   $\text{-}\rho^2$ \\ 
$ \mathbf{6}^{(1)*}$ & 6 &   $6\omega^2$ & $6\omega$ & 0 & 2 &  $2\omega^2$ & $2\omega$ & 0 & 0 & 0 & 0 & 0 &   $\text{-}\omega\sigma$ & $\text{-}\sigma$ & $\text{-}\omega^2\sigma$ &  $\text{-}\sigma^{\ast}$ &  $\text{-}\omega^2\sigma^{\ast}$ &
$\text{-}\omega\sigma^{\ast}$ &    $\text{-}\rho^{2*}$ &    $\text{-}\rho^{4}$ & $\text{-}\rho$ &   $\text{-}\rho^2$ &   $\text{-}\rho^{\ast}$ & $\text{-}\rho^{4*}$ \\
$ \mathbf{6}^{(2)*}$ & 6 & $6\omega^2$ & $6\omega$ & 0 & 2 & $2\omega^2$ & $2\omega$ & 0 & 0 & 0 & 0 & 0 &   $\text{-}\sigma$ & $\text{-}\omega^2\sigma$ &   $\text{-}\omega\sigma$ &  $\text{-}\omega\sigma^{\ast}$ & $\text{-}\sigma^{\ast}$ &  $\text{-}\omega^2\sigma^{\ast}$ &    $\text{-}\rho^{4}$ & $\text{-}\rho$ &    $\text{-}\rho^{2*}$ &   $\text{-}\rho^{4*}$ & $\text{-}\rho^2$ & $\text{-}\rho^{\ast}$\\
\hline
$\mathbf{8}^{(0)}$ & 8 & 8 & 8 & $\text{-}1$ & 0 & 0 & 0 & 0 & 0 & 0 & $\text{-}1$ & $\text{-}1$ & 2 & 2 & 2 & 2 & 2 & 2 & 0 & 0 & 0 & 0 & 0 & 0 \\
$\mathbf{8}^{(1)}$ & 8 & 8 & 8 & $\text{-}1$ & 0 & 0 & 0 & 0 & 0 & 0 & $\text{-}\omega$ & $\text{-}\omega^2$ &    $2\omega$ &    $2\omega$ &    $2\omega$ & $2\omega^2$ &   $2\omega^2$ &   $2\omega^2$ &    0 &    0 &    0 & 0 &    0 &    0 \\
$\mathbf{8}^{(2)}$ & 8 & 8 & 8 & $\text{-}1$ & 0 & 0 & 0 & 0 & 0 & 0 & $\text{-}\omega^2$ & $\text{-}\omega$ &   $2\omega^2$ &   $2\omega^2$ &   $2\omega^2$ & $2\omega$ &    $2\omega$ &    $2\omega$ &    0 &    0 &    0 &    0 & 0 &    0 \\
\hline
$\mathbf{9}$ & 9 &    $9\omega$ &   $9\omega^2$ &   0 & $\text{-}3$ & $\text{-}3\omega$ &  $\text{-}3\omega^2$ & $\text{-}1$ &    $\text{-}\omega$ &   $\text{-}\omega^2$ & 0 & 0 & 0 & 0 & 0 & 0 & 0 & 0 & 0 & 0 & 0 & 0 & 0 & 0 \\
$\mathbf{9}^{\ast}$ & 9 &   $9\omega^2$ &    $9\omega$ &   0 & $\text{-}3$ & $\text{-}3\omega^2$ & $\text{-}3\omega$ & $\text{-}1$ & $\text{-}\omega^2$ & $\text{-}\omega$ & 0 & 0 & 0 & 0 & 0 & 0 & 0 & 0 & 0 & 0 & 0 & 0 & 0 & 0
\\
\end{tabular}
\end{table*}

\begin{table*}[ht!]
\caption{Character table for $\sad$ with $\omega=e^{2\pi i/3}$, $\mu_1=(1-\sqrt{5})/2$, $\mu_2=(1+\sqrt{5})/2$.}
\begin{center}
\begin{tabular}
{c || c | c |c | c  c |c  c | c c | c  c |c | c  c |c | c  c   }

 Size & 1 &72& 90& 45& 45& 72& 72& 120& 120& 90& 90& 72& 72& 72& 45& 1 &1\\
 Ord& 1 & 5 & 4 & 6 & 6 & 15 & 15 & 3 & 3 & 12& 12 & 5 & 15 & 15 & 2& 3 & 3 \\
\hline\hline
$\one^{(0)}$& $1$& $1$ & $1$ & $1$ & $1$ & $1$ & $1$ & $1$ & $1$ & $1$ & $1$ & $1$ & $1$ & $1$ & $1$ & $1$ & $1$ \\\hline
$\three^{(0)}$& $3$& $\mu_2$ & $1$ & -$\omega^*$ & -$\omega$ & $\mu_1\omega^*$ & $\mu_1\omega$ & $0$ & $0$ & $\omega$ & $\omega^*$ & $\mu_1$ & $\mu_2\omega$ & $\mu_2\omega^*$ & -$1$ & $3\omega$ & $3\omega^*$  \\ 
$\three^{(0)*}$& $3$ & $\mu_2$ & $1$ & -$\omega$ & -$\omega^*$ & $\mu_1\omega$ & $\mu_1\omega^*$ & $0$ & $0$ & $\omega^*$ & $\omega$ & $\mu_1$ & $\mu_2\omega^*$ & $\mu_2\omega$ & -$1$ & $3\omega^*$ & $3\omega$ \\
$\three^{(1)}$& $3$ & $\mu_1$ & $1$ & -$\omega^*$ & -$\omega$ & $\mu_2\omega^*$ & $\mu_2\omega$ & $0$ & $0$ & $\omega$ & $\omega^*$ & $\mu_2$ & $\mu_1\omega$ & $\mu_1\omega^*$ & -$1$ & $3\omega$ & $3\omega^*$ \\
$\three^{(1)*}$& $3$ & $\mu_1$ & $1$ & -$\omega$ & -$\omega^*$ & $\mu_2\omega$ & $\mu_2\omega^*$ & $0$ & $0$ & $\omega^*$ & $\omega$ & $\mu_2$ & $\mu_1\omega^*$ & $\mu_1\omega$ & -$1$ & $3\omega^*$ & $3\omega$ \\\hline
$\five^{(0)}$& $5$ & $0$ & -$1$ & $1$ & $1$ & $0$ & $0$ & -$1$ & $2$ & -$1$ & -$1$ & $0$ & $0$ & $0$ & $1$ & $5$ & $5$ \\
$\five^{(0)'}$& $5$ & $0$ & -$1$ & $1$ & $1$ & $0$ & $0$ & $2$ & -$1$ & -$1$ & -$1$ & $0$ & $0$ & $0$ & $1$ & $5$ & $5$ \\\hline
$\six^{(0)}$& $6$ & $1$ & $0$ & $2\omega$ & $2\omega^*$ & $\omega$ & $\omega^*$ & $0$ & $0$ & $0$ & $0$ & $1$ & $\omega^*$ & $\omega$ & $2$ & $6\omega^*$ & $6\omega$ \\
$\six^{(0)*}$& $6$ & $1$ & $0$ & $2\omega^*$ & $2\omega$ & $\omega^*$ & $\omega$ & $0$ & $0$ & $0$ & $0$ & $1$ & $\omega$ & $\omega^*$ & $2$ & $6\omega$ & $6\omega^*$ \\\hline
$\eight^{(0)}$& $8$ & $\mu_2$ & $0$ & $0$ & $0$ & $\mu_1$ & $\mu_1$ & -$1$ & -$1$ & $0$ & $0$ & $\mu_1$ & $\mu_2$ & $\mu_2$ & $0$ & $8$ & $8$ \\
$\eight^{(0)'}$& $8$ & $\mu_1$ & $0$ & $0$ & $0$ & $\mu_2$ & $\mu_2$ & -$1$ & -$1$ & $0$ & $0$ & $\mu_2$ & $\mu_1$ & $\mu_1$ & $0$ & $8$ & $8$ \\\hline
$\nine^{(0)}$& $9$ & -$1$ & $1$ & $1$ & $1$ & -$1$ & -$1$ & $0$ & $0$ & $1$ & $1$ & -$1$ & -$1$ & -$1$ & $1$ & $9$ & $9$ \\
$\nine^{(1)}$& $9$ & -$1$ & $1$ & $\omega$ & $\omega^*$ & -$\omega$ & -$\omega^*$ & $0$ & $0$ & $\omega^*$ & $\omega$ & -$1$ & -$\omega^*$ & -$\omega$ & $1$ & $9\omega^*$ & $9\omega$ \\
$\nine^{(1)*}$& $9$ & -$1$ & $1$ & $\omega^*$ & $\omega$ & -$\omega^*$ & -$\omega$ & $0$ & $0$ & $\omega$ & $\omega^*$ & -$1$ & -$\omega$ & -$\omega^*$ & $1$ & $9\omega$ & $9\omega^*$ \\\hline
$\ten^{(0)}$& $10$ & $0$ & $0$ & -$2$ & -$2$ & $0$ & $0$ & $1$ & $1$ & $0$ & $0$ & $0$ & $0$ & $0$ & -$2$ & $10$ & $10$ \\\hline
$\ften^{(0)}$& $15$ & $0$ & -$1$ & -$\omega^*$ & -$\omega$ & $0$ & $0$ & $0$ & $0$ & -$\omega$ & -$\omega^*$ & $0$ & $0$ & $0$ & -$1$ & $15\omega$ & $15\omega^*$ \\
$\ften^{(0)*}$& $15$ & $0$ & -$1$ & -$\omega$ & -$\omega^*$ & $0$ & $0$ & $0$ & $0$ & -$\omega^*$ & -$\omega$ & $0$ & $0$ & $0$ & -$1$ & $15\omega^*$ & $15\omega$ \\
\end{tabular}
\label{tab: s1080-character-evaluation}
\end{center}
\end{table*}

\section{Basics of Subduction}
\label{sec:subd}

Subduction in group theory refers to the decomposition of an IR of a group into the IRs of one of its subgroups. In the case of $\mathcal{G}$ subducing to $\mathbb{G}$, the infinite set of IRs of $\mathcal{G}$ must be subduced to the finite set of IRs of $\mathbb{G}$. Consequently, only a limited number of IRs can be subduced one-to-one, with higher dimensional IRs subducing to direct sums of increasingly large numbers of IRs. The multiplicity of the $r$-th IR of $\mathbb{G}$ in the subduction of the $j$-th $\mathcal{G}$ IR can be determined by computing
\begin{equation}
\label{eq:sub}
    m_{j}^{(r)} = \frac{1}{g} \sum_{k} n_k \chi_k^{(r)} \chi_{k}^{(j)}
\end{equation}
where $g$ is the size of $\mathbb{G}$, the sum is over all $k$ conjugacy classes,  $n_k$ is the size in $k$ in $\mathbb{G}$, and $\chi_k^{(r)} $ and $ \chi_{k}^{(j)} $ are the characters of the $ r $-th and $ j $-th IRs, respectively, evaluated for elements of the $ k $-th conjugacy class. Using $d_r,d_j$ to indicate the dimension of the IRs, we have
\begin{equation}
\label{eq:drdj}
    d_r=\sum_j  m_{j}^{(r)} d_j
\end{equation}

The requirement that LFT observables must be gauge invariant can be restated as saying they must be functions only of characters of IRs. We therefore anticipate that observables in IRs that subduce to direct sums correspond to larger discrepancies between $\mathcal{G}$ and $\mathbb{G}$. In what follows, we will use the subduction for the IRs of discrete subgroups of $SU(2)$ and $SU(3)$ to interrogate this idea.

\subsection{$SU(2)$ Subduction tables}
By inputting the character tables of $\btt,\bo,\bi$ and Eq.~\ref{eq:su2char} into Eq.~\ref{eq:sub}, we derive the multiplicity of each IR of $\mathbb{G}$ in the subductions of $SU(2)$. 
Due to its relation to the reduction of the continuous spacetime onto the lattice, the subduction of $\bo$ was previously investigated in~\cite{Johnson:1982yq,Moore:2005dw}, and they further consider how this affects states of fixed angular momentum in LGT. It is included alongside $ \btt $ and $\bi$ in Table~\ref{tab:su2subduction}.

Each row of Table~\ref{tab:su2subduction} corresponds to a specific $ SU(2) $ IR, characterized by the quantum number $ j $, with the direct sum of the IRs from the finite subgroups in its subduction. The dimension of the subduced IR must equal the sum of the direct sum. Therefore, with only a finite set of IRs, as $ j $ increase, the number of IRs in the subduction and their multiplicities must grow, with larger discrete groups growing more slowly.

\begin{table*}
    \caption{Subduction of $SU(2)$ to its crystal-like subgroups: $\btt$,$\bo$,$\bi$}
    \label{tab:su2subduction}
    \begin{tabular}{c|lll}
         j&$\btt$  & $\bo$ & $\bi$\\\hline\hline
         0              & $A_0$ & $A_1$ & $A_0$\\
         $\frac{1}{2}$  & $H$ & $G_1$ & $E_1$\\
         1              & $T$ & $T_1$ & $T_1$\\
         $\frac{3}{2}$  & $G_1\oplus G_2$ & $H$ & $G_2$\\
         2              & $E_1\oplus E_2\oplus T$ & $E\oplus T_2$ & $H$\\
         $\frac{5}{2}$  & $H\oplus G_1\oplus G_2$ & $H\oplus G_2$ & $J$\\
         3              & $A_0\oplus 2T$ & $A_2\oplus T_1\oplus T_2$ & $G_1\oplus T_2$\\
         $\frac{7}{2}$  & $2H\oplus G_1\oplus G_2$ & $H\oplus G_1\oplus G_2$ & $J\oplus E_2$\\
         4              & $A_0\oplus E_1\oplus E_2\oplus 2T$ & $E\oplus A_1\oplus T_1\oplus T_2$ & $H\oplus G_1$\\
         $\frac{9}{2}$  & $H\oplus 2G_1\oplus 2G_1$ & $2H\oplus G_1$ & $J\oplus G_2$\\
         5              & $E_1\oplus E_2\oplus 3T$ & $E\oplus 2T_1\oplus T_2$ & $H\oplus T_1\oplus T_2$\\
         $\frac{11}{2}$ & $2H\oplus 2G_1\oplus 2G_2$ & $2H\oplus G_1\oplus G_2$ & $J\oplus E_1\oplus G_2$\\
         6              & $2A_0\oplus E_1\oplus E_2\oplus 3T$ & $E\oplus A_1\oplus A_2\oplus T_1\oplus 2T_2$ & $H\oplus A_0\oplus G_1\oplus T_1$\\
         $\frac{13}{2}$ & $3H\oplus 2G_1\oplus 2G_2$ & $2H\oplus G_1\oplus 2G_2$ & $J\oplus E_1\oplus E_2\oplus G_2$\\
         7              & $A_0\oplus E_1\oplus E_2\oplus 4T$ & $E\oplus A_2\oplus 2T_1\oplus 2T_2$ & $H\oplus G_1\oplus T_1\oplus T_2$\\
         $\frac{15}{2}$ & $2H\oplus 3G_1\oplus 3G_2$ & $3H\oplus G_1\oplus G_2$ & $2J\oplus E_2\oplus G_2$\\
         8              & $A_0\oplus 2E_1\oplus 2E_2\oplus 4T$ & $2E\oplus A_1\oplus 2T_1\oplus 2T_2$ & $2H\oplus H_1\oplus T_2$\\
         $\frac{17}{2}$ & $3H\oplus 3G_1\oplus 3G_2$ & $3H\oplus 2G_1\oplus G_2$ & $2J\oplus E_2\oplus G_2$\\
         9              & $2A_0\oplus E_1\oplus E_2\oplus 5T$ & $E\oplus A_1\oplus A_2\oplus 3T_1\oplus 2T_2$ & $H\oplus 2G_1\oplus T_1\oplus T_2$\\
         $\frac{19}{2}$ & $4H\oplus 3G_1\oplus 3G_2$ & $3H\oplus 2G_1\oplus 2G_2$ & $2J\oplus E_1\oplus E_2\oplus G_2$\\
         10             & $2A_0\oplus 2E_1\oplus 2E_2\oplus 5T$ & $2E\oplus A_1\oplus A_2\oplus 2T_1\oplus 3T_2$ & $2H\oplus A_0\oplus G_1\oplus T_1\oplus T_2$\\
    \end{tabular}
\end{table*}

The table shows that all three crystal-like subgroups have one-to-one subductions of the lowest IRs of $SU(2)$, increasing from 3 for $\btt$ to 4 for $\bo$ to 6 for $ \bi $.  While the gauge-invariant spectrum does not map precisely to the increasing dimension of IRs, it is perhaps instructive to consider the similar example of rigid rotators~\cite{Johnson:1982yq,Moore:2005dw} where the energy levels scale like $j^2$. Taking this heuristic, the digitization approximation might be estimated to break down for the first $j$ for which the subduction is not one-to-one.  This would suggest that the relative breakdown of $\btt$ to $\bo$ and $\bi$ would be 1:1.8:4. From Euclidean simulations in 2+1d and 3+1d, it has been observed that the ratio of freezeout couplings for the three groups are 1:1.6:2.8 and 1:1.4:2.6 respectively~\cite{Gustafson:2022xdt,Gustafson:2023kvd}. Thus, the subduction to multiple IRs does not seem to be more than qualitatively related to the freezeout.

Similar to the question asked by~\cite{Johnson:1982yq,Moore:2005dw}:  what is the spin content of a lattice eigenstate that transforms as a single IR, we are interested in what IRs of the continuous group are mixed in the discrete group.  From this, one could explore the hierarchy of states to motivate which higher IRs should be added to the interactions to reduce the effects of digitization. As such, we list here the induced IRs up to $j=10$ for each crystal-like group.  To provide the reader with a sense of the potential importance of different IRs of $\mathbb{G}$, we have listed their dimensionality $d$ as well.  An IR for which $h$ multiple copies are required in the induction is denoted by $|_h$.  

For the 7 IRs of $\btt$, we observe the number of IRs of $SU(2)$ which mix into $\btt$ IRs are
\begin{align}
       (d=1)\;& A_0\rightarrow  (j=0,3,4,6|_2,7,8,9|_2,10,\hdots)\notag\\
       (d=1)\;& E_1,E_2\rightarrow  (j=2,4,5,6,7,8|_2,9,10|_2,\hdots)\notag\\
       (d=2)\;& H\rightarrow  \textstyle (j=\frac{1}{2},\frac{5}{2},\frac{7}{2}|_2,\frac{9}{2},\frac{11}{2}|_2,\frac{13}{2}|_3,\notag\\&\phantom{xxxxxxxxxxxx} \textstyle\frac{15}{2}|_2,\frac{17}{2}|_3,\frac{19}{2}|_4,\hdots)\notag\\
       (d=2)\;& G_1,G_2\rightarrow  \textstyle(j=\frac{3}{2},\frac{5}{2},\frac{7}{2},\frac{9}{2}|_2,\frac{11}{2}|_2,\frac{13}{2}|_2,\notag\\&\phantom{xxxxxxxxxxxx} \textstyle\frac{15}{2}|_3,\frac{17}{2}|_3,\frac{19}{2}|_3,\hdots)\notag\\
       (d=3)\;& T\rightarrow  \textstyle(j=1,2,3|_2,4|_2,5|_3,6|_3,7|_4,8|_4,\notag\\&\phantom{xxxxxxxxxxxx} \textstyle9|_5,10|_5,\hdots)\notag
\end{align}
Because $SU(2)$ has one 1d IR, we see that $E_1,E_2$ arise only within direct sums of IRs.  Further, they always occur as a pair.  A similar behavior is observed in the 2d IRs of $\btt$.  One of them is subduced from the $j=1/2$ IR of $SU(2)$, while the other two only occur in pairs, first together in the $j=3/2$ IR. 

Given that $\btt$ is a subgroup of $\bo$, it is useful to compare their IRs. In contrast to $\btt$, $\bo$ has only two 1d IRs, an additional 3d one, and a new 4d IR that is the induction of $G_1,G_2$ of $\btt$. The set of $\bo$ IRs consist of a different mixing of $SU(2)$ IRs, with generically greater separation in $d_j$:
\begin{align}
(d=1)\;&A_1\rightarrow (j=0,4,6,8,9,10,\hdots)\notag\\          (d=1)\;&A_2\rightarrow (j=3,6,7,9,10,\hdots)\notag\\
         (d=2)\;&E\rightarrow (j=2,4,5,6,7,8|_2,9,10|_2,\hdots)\notag\\
         (d=2)\;&G_1\rightarrow \textstyle (j=\frac{1}{2},\frac{7}{2},\frac{9}{2},\frac{11}{2},\frac{13}{2},\frac{15}{2},\frac{17}{2}|_2,\frac{19}{2}|_2,\hdots)\notag\\
         (d=2)\;&G_2\rightarrow \textstyle (j=\frac{5}{2},\frac{7}{2},\frac{11}{2},\frac{13}{2}|_2,\frac{15}{2},\frac{17}{2}|_2,\frac{19}{2}|_2,\hdots)\notag\\
         (d=3)\;&T_1\rightarrow \textstyle (j=1,3,4,5|_2,6,7|_2,8|_2,9|_3,10|_2,\hdots)\notag\\
         (d=3)\;&T_2\rightarrow \textstyle (j=2,3,4,5,6|_2,7|_2,8|_2,9|_2,10|_3,\hdots)\notag\\
         (d=4)\;&H\rightarrow \textstyle (j=\frac{3}{2},\frac{5}{2},\frac{7}{2},\frac{9}{2}|_2,\frac{11}{2}|_2,\frac{13}{2}|_2,\notag\\&\phantom{xxxxxxxxxxxxxxx} \textstyle\frac{15}{2}|_3,\frac{17}{2}|_3,\frac{19}{2}|_3,\hdots)\notag
\end{align}
By inspection, we see that, unlike $\btt$, no pairs of IRs exist in the induction.  Concerning the fundamental faithful IR used in the Wilson action, $G_1$, we note that while it does induce to $j=1/2$ like $H$ for $\btt$, it mixes with the higher dimension $j=7/2$ rather than $5/2$ for $H$.  This should be related to a larger freezeout coupling and thus a better approximation of $SU(2)$.

Finally, we can investigate $\bi$, observing that for each $d_j$, one IR corresponds one-to-one with $SU(2)$, and the others occur in complex mixings. The inductions of $\bi$ are:
\begin{align}
         (d=1)\;&A_0\rightarrow \textstyle (j=0,6,10\hdots)\notag\\
         (d=2)\;&E_1\rightarrow \textstyle (j=\frac{1}{2},\frac{11}{2},\frac{13}{2},\frac{19}{2},\hdots)\notag\\
         (d=2)\;&E_2\rightarrow \textstyle (j=\frac{7}{2},\frac{13}{2},\frac{17}{2},\frac{19}{2}\hdots)\notag\\
         (d=3)\;&T_1\rightarrow \textstyle (j=1,5,6,7,9,10\hdots)\notag\\
         (d=3)\;&T_2\rightarrow \textstyle (j=3,5,7,8,9,10\hdots)\notag\\
         (d=4)\;&G_1\rightarrow \textstyle (j=3,4,6,7,8,9|_2,10,\hdots)\notag\\
         (d=4)\;&G_2\rightarrow \textstyle (j=\frac{3}{2},\frac{9}{2},\frac{11}{2},\frac{13}{2},\frac{15}{2},\frac{17}{2},\frac{19}{2},\hdots)\notag\\
         (d=5)\;&H\rightarrow \textstyle (j=2,4,5,6,7,8|_2,9,10|_2\hdots)\notag\\
         (d=6)\;&J\rightarrow \textstyle (j=\frac{5}{2},\frac{7}{2},\frac{9}{2},\frac{11}{2},\frac{13}{2},\frac{15}{2}|_2,\frac{17}{2}|_2,\frac{19}{2}|_2,\hdots)\notag
\end{align}

Looking again at the 2d IR used in the Wilson action, $E_1$, we see the next state after $j=1/2$ to mix into it has increased further to $j=11/2$. Previous results using group space decimation suggested in~\cite{Ji:2020kjk} would suggest and IR enter as $\beta^{d_j-1}$ in the strong-coupling expansion and thus the subduction gives a relation to breaking at strong coupling. A dedicated study of excited state glueball masses should be undertaken to study how the dynamics of strong-coupling eigenstates of a given IR of a subgroup are affected by their mixing into multiple IRs of $SU(2)$, especially when $m_{j}^{r}>1$. 

\subsection{$SU(3)$ Subduction tables}
\label{ssec:subtab}

\begin{table*}
    \caption{Subduction of $SU(3)$ to crystal-like subgroups: $\Sigma(108)$ , $\Sigma(216)$ , $\Sigma(648)$, $\Sigma(1080)$}
    \label{tab:su3subduction}
\scalebox{0.6}{
    \begin{tabular}{c|llll}
        $(p,q)$ & $\Sigma(108)$ & $\Sigma(216)$  & $\Sigma(648)$ & $\Sigma(1080)$\\\hline\hline
(0,0)&$	\mathbf{1}^{(0)}$ & $	\mathbf{1}^{(0)}$ & $	\mathbf{1}^{(0)}$ & $	\mathbf{1}^{(0)}$ \\ 
(1,0)&$	\mathbf{3}^{(0)*}$ & $	\mathbf{3}^{(0)*}$ & $	\mathbf{3}^{(0)*}$ & $	\mathbf{3}^{(0)*}$ \\ 
(0,1)&$	\mathbf{3}^{(0)}$ & $	\mathbf{3}^{(0)}$ & $	\mathbf{3}^{(0)}$ & $	\mathbf{3}^{(0)}$ \\ 
(2,0)&$	\mathbf{3}^{(1)}\oplus \mathbf{3}^{(3)}$ & $	\mathbf{6}^{(0)}$ & $	\mathbf{6}^{(0)}$ & $	\mathbf{6}^{(0)*}$ \\ 
(1,1)&$	\mathbf{4}^{(0)}\oplus \mathbf{4}^{(0)'}$ & $	\mathbf{8}^{(0)}$ & $	\mathbf{8}^{(0)}$ & $	\mathbf{8}^{(0)}$ \\ 
(0,2)&$	\mathbf{3}^{(1)*}\oplus \mathbf{3}^{(3)*}$ & $	\mathbf{6}^{(0)*}$ & $	\mathbf{6}^{(0)*}$ & $	\mathbf{6}^{(0)}$ \\ 
(3,0)&$	\mathbf{1}^{(1)}\oplus \mathbf{1}^{(3)}\oplus \mathbf{4}^{(0)}\oplus \mathbf{4}^{(0)'}$ & $	\mathbf{2}^{(0)}\oplus \mathbf{8}^{(0)}$ & $	\mathbf{2}^{(2)}\oplus \mathbf{8}^{(1)}$ & $	\mathbf{10}^{(0)}$ \\ 
(2,1)&$	\mathbf{3}^{(0)*}\oplus \mathbf{3}^{(1)*}\oplus \mathbf{3}^{(3)*}\oplus 2\,\mathbf{3}^{(2)*}$ & $	\mathbf{3}^{(1)*}\oplus \mathbf{3}^{(2)*}\oplus \mathbf{3}^{(3)*}\oplus \mathbf{6}^{(0)*}$ & $	\mathbf{6}^{(2)*}\oplus \mathbf{9}^{(0)*}$ & $	\mathbf{15}^{(0)*}$ \\ 
(1,2)&$	\mathbf{3}^{(0)}\oplus \mathbf{3}^{(1)}\oplus \mathbf{3}^{(3)}\oplus 2\,\mathbf{3}^{(2)}$ & $	\mathbf{3}^{(1)}\oplus \mathbf{3}^{(2)}\oplus \mathbf{3}^{(3)}\oplus \mathbf{6}^{(0)}$ & $	\mathbf{6}^{(2)}\oplus \mathbf{9}^{(0)}$ & $	\mathbf{15}^{(0)}$ \\ 
(0,3)&$	\mathbf{1}^{(1)}\oplus \mathbf{1}^{(3)}\oplus \mathbf{4}^{(0)}\oplus \mathbf{4}^{(0)'}$ & $	\mathbf{2}^{(0)}\oplus \mathbf{8}^{(0)}$ & $	\mathbf{2}^{(1)}\oplus \mathbf{8}^{(2)}$ & $	\mathbf{10}^{(0)}$ \\ 
(4,0)&$	\mathbf{3}^{(0)*}\oplus 2\,\mathbf{3}^{(1)*}\oplus 2\,\mathbf{3}^{(3)*}$ & $	\mathbf{3}^{(0)*}\oplus 2\,\mathbf{6}^{(0)*}$ & $	\mathbf{3}^{(2)*}\oplus \mathbf{6}^{(1)*}\oplus \mathbf{6}^{(2)*}$ & $	\mathbf{6}^{(0)}\oplus \mathbf{9}^{(1)}$ \\ 
(3,1)&$	2\,\mathbf{3}^{(0)}\oplus 2\,\mathbf{3}^{(1)}\oplus 2\,\mathbf{3}^{(2)}\oplus 2\,\mathbf{3}^{(3)}$ & $	\mathbf{3}^{(0)}\oplus \mathbf{3}^{(1)}\oplus \mathbf{3}^{(2)}\oplus \mathbf{3}^{(3)}\oplus 2\,\mathbf{6}^{(0)}$ & $	\mathbf{3}^{(1)}\oplus \mathbf{6}^{(0)}\oplus \mathbf{6}^{(1)}\oplus \mathbf{9}^{(0)}$ & $	\mathbf{15}^{(0)}\oplus \mathbf{9}^{(1)*}$ \\ 
(2,2)&$	\mathbf{1}^{(0)}\oplus 2\,\mathbf{1}^{(2)}\oplus 3\,\mathbf{4}^{(0)}\oplus 3\,\mathbf{4}^{(0)'}$ & $	\mathbf{1}^{(1)}\oplus \mathbf{1}^{(2)}\oplus \mathbf{1}^{(3)}\oplus 3\,\mathbf{8}^{(0)}$ & $	\mathbf{3}^{(a)}\oplus \mathbf{8}^{(0)}\oplus \mathbf{8}^{(1)}\oplus \mathbf{8}^{(2)}$ & $	\mathbf{5}^{(0)'}\oplus \mathbf{5}^{(0)}\oplus \mathbf{8}^{(0)'}\oplus \mathbf{9}^{(0)}$ \\ 
(1,3)&$	2\,\mathbf{3}^{(0)*}\oplus 2\,\mathbf{3}^{(1)*}\oplus 2\,\mathbf{3}^{(2)*}\oplus 2\,\mathbf{3}^{(3)*}$ & $	\mathbf{3}^{(0)*}\oplus \mathbf{3}^{(1)*}\oplus \mathbf{3}^{(2)*}\oplus \mathbf{3}^{(3)*}\oplus 2\,\mathbf{6}^{(0)*}$ & $	\mathbf{3}^{(1)*}\oplus \mathbf{6}^{(0)*}\oplus \mathbf{6}^{(1)*}\oplus \mathbf{9}^{(0)*}$ & $	\mathbf{15}^{(0)*}\oplus \mathbf{9}^{(1)}$ \\ 
(0,4)&$	\mathbf{3}^{(0)}\oplus 2\,\mathbf{3}^{(1)}\oplus 2\,\mathbf{3}^{(3)}$ & $	\mathbf{3}^{(0)}\oplus 2\,\mathbf{6}^{(0)}$ & $	\mathbf{3}^{(2)}\oplus \mathbf{6}^{(1)}\oplus \mathbf{6}^{(2)}$ & $	\mathbf{6}^{(0)*}\oplus \mathbf{9}^{(1)*}$ \\ 
(5,0)&$	\mathbf{3}^{(1)}\oplus \mathbf{3}^{(3)}\oplus 2\,\mathbf{3}^{(2)}\oplus 3\,\mathbf{3}^{(0)}$ & $	\mathbf{3}^{(1)}\oplus \mathbf{3}^{(2)}\oplus \mathbf{3}^{(3)}\oplus \mathbf{6}^{(0)}\oplus 2\,\mathbf{3}^{(0)}$ & $	\mathbf{3}^{(1)}\oplus \mathbf{3}^{(2)}\oplus \mathbf{6}^{(1)}\oplus \mathbf{9}^{(0)}$ & $	\mathbf{15}^{(0)}\oplus \mathbf{3}^{(0)}\oplus \mathbf{3}^{(1)}$ \\ 
(4,1)&$	\mathbf{1}^{(0)}\oplus \mathbf{1}^{(1)}\oplus \mathbf{1}^{(3)}\oplus 4\,\mathbf{4}^{(0)}\oplus 4\,\mathbf{4}^{(0)'}$ & $	\mathbf{1}^{(0)}\oplus \mathbf{2}^{(0)}\oplus 4\,\mathbf{8}^{(0)}$ & $	\mathbf{1}^{(1)}\oplus \mathbf{2}^{(0)}\oplus \mathbf{8}^{(0)}\oplus \mathbf{8}^{(2)}\oplus 2\,\mathbf{8}^{(1)}$ & $	\mathbf{10}^{(0)}\oplus \mathbf{8}^{(0)'}\oplus \mathbf{8}^{(0)}\oplus \mathbf{9}^{(0)}$ \\ 
(3,2)&$	3\,\mathbf{3}^{(1)*}\oplus 3\,\mathbf{3}^{(3)*}\oplus 4\,\mathbf{3}^{(0)*}\oplus 4\,\mathbf{3}^{(2)*}$ & $	2\,\mathbf{3}^{(0)*}\oplus 2\,\mathbf{3}^{(1)*}\oplus 2\,\mathbf{3}^{(2)*}\oplus 2\,\mathbf{3}^{(3)*}\oplus 3\,\mathbf{6}^{(0)*}$ & $	\mathbf{3}^{(0)*}\oplus \mathbf{3}^{(2)*}\oplus \mathbf{6}^{(0)*}\oplus \mathbf{6}^{(1)*}\oplus \mathbf{6}^{(2)*}\oplus 2\,\mathbf{9}^{(0)*}$ & $	\mathbf{3}^{(1)*}\oplus \mathbf{9}^{(1)}\oplus 2\,\mathbf{15}^{(0)*}$ \\ 
(2,3)&$	3\,\mathbf{3}^{(1)}\oplus 3\,\mathbf{3}^{(3)}\oplus 4\,\mathbf{3}^{(0)}\oplus 4\,\mathbf{3}^{(2)}$ & $	2\,\mathbf{3}^{(0)}\oplus 2\,\mathbf{3}^{(1)}\oplus 2\,\mathbf{3}^{(2)}\oplus 2\,\mathbf{3}^{(3)}\oplus 3\,\mathbf{6}^{(0)}$ & $	\mathbf{3}^{(0)}\oplus \mathbf{3}^{(2)}\oplus \mathbf{6}^{(0)}\oplus \mathbf{6}^{(1)}\oplus \mathbf{6}^{(2)}\oplus 2\,\mathbf{9}^{(0)}$ & $	\mathbf{3}^{(1)}\oplus \mathbf{9}^{(1)*}\oplus 2\,\mathbf{15}^{(0)}$ \\ 
(1,4)&$	\mathbf{1}^{(0)}\oplus \mathbf{1}^{(1)}\oplus \mathbf{1}^{(3)}\oplus 4\,\mathbf{4}^{(0)}\oplus 4\,\mathbf{4}^{(0)'}$ & $	\mathbf{1}^{(0)}\oplus \mathbf{2}^{(0)}\oplus 4\,\mathbf{8}^{(0)}$ & $	\mathbf{1}^{(2)}\oplus \mathbf{2}^{(0)}\oplus \mathbf{8}^{(0)}\oplus \mathbf{8}^{(1)}\oplus 2\,\mathbf{8}^{(2)}$ & $	\mathbf{10}^{(0)}\oplus \mathbf{8}^{(0)'}\oplus \mathbf{8}^{(0)}\oplus \mathbf{9}^{(0)}$ \\ 
(0,5)&$	\mathbf{3}^{(1)*}\oplus \mathbf{3}^{(3)*}\oplus 2\,\mathbf{3}^{(2)*}\oplus 3\,\mathbf{3}^{(0)*}$ & $	\mathbf{3}^{(1)*}\oplus \mathbf{3}^{(2)*}\oplus \mathbf{3}^{(3)*}\oplus \mathbf{6}^{(0)*}\oplus 2\,\mathbf{3}^{(0)*}$ & $	\mathbf{3}^{(1)*}\oplus \mathbf{3}^{(2)*}\oplus \mathbf{6}^{(1)*}\oplus \mathbf{9}^{(0)*}$ & $	\mathbf{15}^{(0)*}\oplus \mathbf{3}^{(0)*}\oplus \mathbf{3}^{(1)*}$ \\ 
(6,0)&$	2\,\mathbf{1}^{(0)}\oplus 2\,\mathbf{1}^{(2)}\oplus 3\,\mathbf{4}^{(0)}\oplus 3\,\mathbf{4}^{(0)'}$ & $	\mathbf{1}^{(0)}\oplus \mathbf{1}^{(1)}\oplus \mathbf{1}^{(2)}\oplus \mathbf{1}^{(3)}\oplus 3\,\mathbf{8}^{(0)}$ & $	\mathbf{1}^{(2)}\oplus \mathbf{3}^{(a)}\oplus \mathbf{8}^{(1)}\oplus 2\,\mathbf{8}^{(2)}$ & $	\mathbf{1}^{(0)}\oplus \mathbf{5}^{(0)'}\oplus \mathbf{5}^{(0)}\oplus \mathbf{8}^{(0)}\oplus \mathbf{9}^{(0)}$ \\ 
(5,1)&$	4\,\mathbf{3}^{(0)*}\oplus 4\,\mathbf{3}^{(1)*}\oplus 4\,\mathbf{3}^{(2)*}\oplus 4\,\mathbf{3}^{(3)*}$ & $	2\,\mathbf{3}^{(0)*}\oplus 2\,\mathbf{3}^{(1)*}\oplus 2\,\mathbf{3}^{(2)*}\oplus 2\,\mathbf{3}^{(3)*}\oplus 4\,\mathbf{6}^{(0)*}$ & $	\mathbf{3}^{(1)*}\oplus \mathbf{3}^{(2)*}\oplus \mathbf{6}^{(0)*}\oplus \mathbf{6}^{(2)*}\oplus 2\,\mathbf{6}^{(1)*}\oplus 2\,\mathbf{9}^{(0)*}$ & $	\mathbf{3}^{(0)*}\oplus \mathbf{6}^{(0)}\oplus \mathbf{9}^{(1)}\oplus 2\,\mathbf{15}^{(0)*}$ \\ 
(4,2)&$	4\,\mathbf{3}^{(0)}\oplus 4\,\mathbf{3}^{(2)}\oplus 6\,\mathbf{3}^{(1)}\oplus 6\,\mathbf{3}^{(3)}$ & $	2\,\mathbf{3}^{(0)}\oplus 2\,\mathbf{3}^{(1)}\oplus 2\,\mathbf{3}^{(2)}\oplus 2\,\mathbf{3}^{(3)}\oplus 6\,\mathbf{6}^{(0)}$ & $	\mathbf{3}^{(0)}\oplus \mathbf{3}^{(1)}\oplus 2\,\mathbf{6}^{(0)}\oplus 2\,\mathbf{6}^{(1)}\oplus 2\,\mathbf{6}^{(2)}\oplus 2\,\mathbf{9}^{(0)}$ & $	2\,\mathbf{15}^{(0)}\oplus 2\,\mathbf{6}^{(0)*}\oplus 2\,\mathbf{9}^{(1)*}$ \\ 
(3,3)&$	2\,\mathbf{1}^{(0)}\oplus 2\,\mathbf{1}^{(1)}\oplus 2\,\mathbf{1}^{(2)}\oplus 2\,\mathbf{1}^{(3)}\oplus 7\,\mathbf{4}^{(0)}\oplus 7\,\mathbf{4}^{(0)'}$ & $	\mathbf{1}^{(0)}\oplus \mathbf{1}^{(1)}\oplus \mathbf{1}^{(2)}\oplus \mathbf{1}^{(3)}\oplus 2\,\mathbf{2}^{(0)}\oplus 7\,\mathbf{8}^{(0)}$ & $	\mathbf{1}^{(0)}\oplus \mathbf{2}^{(1)}\oplus \mathbf{2}^{(2)}\oplus \mathbf{3}^{(a)}\oplus 2\,\mathbf{8}^{(1)}\oplus 2\,\mathbf{8}^{(2)}\oplus 3\,\mathbf{8}^{(0)}$ & $	\mathbf{5}^{(0)'}\oplus \mathbf{5}^{(0)}\oplus \mathbf{8}^{(0)'}\oplus \mathbf{8}^{(0)}\oplus 2\,\mathbf{10}^{(0)}\oplus 2\,\mathbf{9}^{(0)}$ \\ 
(2,4)&$	4\,\mathbf{3}^{(0)*}\oplus 4\,\mathbf{3}^{(2)*}\oplus 6\,\mathbf{3}^{(1)*}\oplus 6\,\mathbf{3}^{(3)*}$ & $	2\,\mathbf{3}^{(0)*}\oplus 2\,\mathbf{3}^{(1)*}\oplus 2\,\mathbf{3}^{(2)*}\oplus 2\,\mathbf{3}^{(3)*}\oplus 6\,\mathbf{6}^{(0)*}$ & $	\mathbf{3}^{(0)*}\oplus \mathbf{3}^{(1)*}\oplus 2\,\mathbf{6}^{(0)*}\oplus 2\,\mathbf{6}^{(1)*}\oplus 2\,\mathbf{6}^{(2)*}\oplus 2\,\mathbf{9}^{(0)*}$ & $	2\,\mathbf{15}^{(0)*}\oplus 2\,\mathbf{6}^{(0)}\oplus 2\,\mathbf{9}^{(1)}$ \\ 
(1,5)&$	4\,\mathbf{3}^{(0)}\oplus 4\,\mathbf{3}^{(1)}\oplus 4\,\mathbf{3}^{(2)}\oplus 4\,\mathbf{3}^{(3)}$ & $	2\,\mathbf{3}^{(0)}\oplus 2\,\mathbf{3}^{(1)}\oplus 2\,\mathbf{3}^{(2)}\oplus 2\,\mathbf{3}^{(3)}\oplus 4\,\mathbf{6}^{(0)}$ & $	\mathbf{3}^{(1)}\oplus \mathbf{3}^{(2)}\oplus \mathbf{6}^{(0)}\oplus \mathbf{6}^{(2)}\oplus 2\,\mathbf{6}^{(1)}\oplus 2\,\mathbf{9}^{(0)}$ & $	\mathbf{3}^{(0)}\oplus \mathbf{6}^{(0)*}\oplus \mathbf{9}^{(1)*}\oplus 2\,\mathbf{15}^{(0)}$ \\ 
(0,6)&$	2\,\mathbf{1}^{(0)}\oplus 2\,\mathbf{1}^{(2)}\oplus 3\,\mathbf{4}^{(0)}\oplus 3\,\mathbf{4}^{(0)'}$ & $	\mathbf{1}^{(0)}\oplus \mathbf{1}^{(1)}\oplus \mathbf{1}^{(2)}\oplus \mathbf{1}^{(3)}\oplus 3\,\mathbf{8}^{(0)}$ & $	\mathbf{1}^{(1)}\oplus \mathbf{3}^{(a)}\oplus \mathbf{8}^{(2)}\oplus 2\,\mathbf{8}^{(1)}$ & $	\mathbf{1}^{(0)}\oplus \mathbf{5}^{(0)'}\oplus \mathbf{5}^{(0)}\oplus \mathbf{8}^{(0)}\oplus \mathbf{9}^{(0)}$ \\ 
(7,0)&$	2\,\mathbf{3}^{(1)*}\oplus 2\,\mathbf{3}^{(3)*}\oplus 4\,\mathbf{3}^{(0)*}\oplus 4\,\mathbf{3}^{(2)*}$ & $	2\,\mathbf{3}^{(0)*}\oplus 2\,\mathbf{3}^{(1)*}\oplus 2\,\mathbf{3}^{(2)*}\oplus 2\,\mathbf{3}^{(3)*}\oplus 2\,\mathbf{6}^{(0)*}$ & $	\mathbf{6}^{(0)*}\oplus \mathbf{6}^{(1)*}\oplus 2\,\mathbf{3}^{(1)*}\oplus 2\,\mathbf{9}^{(0)*}$ & $	\mathbf{3}^{(0)*}\oplus \mathbf{3}^{(1)*}\oplus 2\,\mathbf{15}^{(0)*}$ \\ 
(6,1)&$	5\,\mathbf{3}^{(0)}\oplus 5\,\mathbf{3}^{(1)}\oplus 5\,\mathbf{3}^{(3)}\oplus 6\,\mathbf{3}^{(2)}$ & $	2\,\mathbf{3}^{(0)}\oplus 3\,\mathbf{3}^{(1)}\oplus 3\,\mathbf{3}^{(2)}\oplus 3\,\mathbf{3}^{(3)}\oplus 5\,\mathbf{6}^{(0)}$ & $	\mathbf{6}^{(0)}\oplus 2\,\mathbf{3}^{(2)}\oplus 2\,\mathbf{6}^{(1)}\oplus 2\,\mathbf{6}^{(2)}\oplus 3\,\mathbf{9}^{(0)}$ & $	\mathbf{3}^{(0)}\oplus \mathbf{6}^{(0)*}\oplus \mathbf{9}^{(1)*}\oplus 3\,\mathbf{15}^{(0)}$ \\ 
(5,2)&$	\mathbf{1}^{(0)}\oplus 2\,\mathbf{1}^{(2)}\oplus 3\,\mathbf{1}^{(1)}\oplus 3\,\mathbf{1}^{(3)}\oplus 9\,\mathbf{4}^{(0)}\oplus 9\,\mathbf{4}^{(0)'}$ & $	\mathbf{1}^{(1)}\oplus \mathbf{1}^{(2)}\oplus \mathbf{1}^{(3)}\oplus 3\,\mathbf{2}^{(0)}\oplus 9\,\mathbf{8}^{(0)}$ & $	\mathbf{2}^{(0)}\oplus \mathbf{2}^{(1)}\oplus \mathbf{2}^{(2)}\oplus \mathbf{3}^{(a)}\oplus 3\,\mathbf{8}^{(0)}\oplus 3\,\mathbf{8}^{(1)}\oplus 3\,\mathbf{8}^{(2)}$ & $	\mathbf{5}^{(0)'}\oplus \mathbf{5}^{(0)}\oplus \mathbf{9}^{(0)}\oplus 2\,\mathbf{8}^{(0)'}\oplus 2\,\mathbf{8}^{(0)}\oplus 3\,\mathbf{10}^{(0)}$ \\ 
(4,3)&$	7\,\mathbf{3}^{(1)*}\oplus 7\,\mathbf{3}^{(3)*}\oplus 8\,\mathbf{3}^{(0)*}\oplus 8\,\mathbf{3}^{(2)*}$ & $	4\,\mathbf{3}^{(0)*}\oplus 4\,\mathbf{3}^{(1)*}\oplus 4\,\mathbf{3}^{(2)*}\oplus 4\,\mathbf{3}^{(3)*}\oplus 7\,\mathbf{6}^{(0)*}$ & $	\mathbf{3}^{(1)*}\oplus \mathbf{3}^{(2)*}\oplus \mathbf{6}^{(1)*}\oplus 2\,\mathbf{3}^{(0)*}\oplus 3\,\mathbf{6}^{(0)*}\oplus 3\,\mathbf{6}^{(2)*}\oplus 4\,\mathbf{9}^{(0)*}$ & $	\mathbf{3}^{(0)*}\oplus \mathbf{3}^{(1)*}\oplus \mathbf{6}^{(0)}\oplus 2\,\mathbf{9}^{(1)}\oplus 4\,\mathbf{15}^{(0)*}$ \\ 
(3,4)&$	7\,\mathbf{3}^{(1)}\oplus 7\,\mathbf{3}^{(3)}\oplus 8\,\mathbf{3}^{(0)}\oplus 8\,\mathbf{3}^{(2)}$ & $	4\,\mathbf{3}^{(0)}\oplus 4\,\mathbf{3}^{(1)}\oplus 4\,\mathbf{3}^{(2)}\oplus 4\,\mathbf{3}^{(3)}\oplus 7\,\mathbf{6}^{(0)}$ & $	\mathbf{3}^{(1)}\oplus \mathbf{3}^{(2)}\oplus \mathbf{6}^{(1)}\oplus 2\,\mathbf{3}^{(0)}\oplus 3\,\mathbf{6}^{(0)}\oplus 3\,\mathbf{6}^{(2)}\oplus 4\,\mathbf{9}^{(0)}$ & $	\mathbf{3}^{(0)}\oplus \mathbf{3}^{(1)}\oplus \mathbf{6}^{(0)*}\oplus 2\,\mathbf{9}^{(1)*}\oplus 4\,\mathbf{15}^{(0)}$ \\ 
(2,5)&$	\mathbf{1}^{(0)}\oplus 2\,\mathbf{1}^{(2)}\oplus 3\,\mathbf{1}^{(1)}\oplus 3\,\mathbf{1}^{(3)}\oplus 9\,\mathbf{4}^{(0)}\oplus 9\,\mathbf{4}^{(0)'}$ & $	\mathbf{1}^{(1)}\oplus \mathbf{1}^{(2)}\oplus \mathbf{1}^{(3)}\oplus 3\,\mathbf{2}^{(0)}\oplus 9\,\mathbf{8}^{(0)}$ & $	\mathbf{2}^{(0)}\oplus \mathbf{2}^{(1)}\oplus \mathbf{2}^{(2)}\oplus \mathbf{3}^{(a)}\oplus 3\,\mathbf{8}^{(0)}\oplus 3\,\mathbf{8}^{(1)}\oplus 3\,\mathbf{8}^{(2)}$ & $	\mathbf{5}^{(0)'}\oplus \mathbf{5}^{(0)}\oplus \mathbf{9}^{(0)}\oplus 2\,\mathbf{8}^{(0)'}\oplus 2\,\mathbf{8}^{(0)}\oplus 3\,\mathbf{10}^{(0)}$ \\ 
(1,6)&$	5\,\mathbf{3}^{(0)*}\oplus 5\,\mathbf{3}^{(1)*}\oplus 5\,\mathbf{3}^{(3)*}\oplus 6\,\mathbf{3}^{(2)*}$ & $	2\,\mathbf{3}^{(0)*}\oplus 3\,\mathbf{3}^{(1)*}\oplus 3\,\mathbf{3}^{(2)*}\oplus 3\,\mathbf{3}^{(3)*}\oplus 5\,\mathbf{6}^{(0)*}$ & $	\mathbf{6}^{(0)*}\oplus 2\,\mathbf{3}^{(2)*}\oplus 2\,\mathbf{6}^{(1)*}\oplus 2\,\mathbf{6}^{(2)*}\oplus 3\,\mathbf{9}^{(0)*}$ & $	\mathbf{3}^{(0)*}\oplus \mathbf{6}^{(0)}\oplus \mathbf{9}^{(1)}\oplus 3\,\mathbf{15}^{(0)*}$ \\ 
(0,7)&$	2\,\mathbf{3}^{(1)}\oplus 2\,\mathbf{3}^{(3)}\oplus 4\,\mathbf{3}^{(0)}\oplus 4\,\mathbf{3}^{(2)}$ & $	2\,\mathbf{3}^{(0)}\oplus 2\,\mathbf{3}^{(1)}\oplus 2\,\mathbf{3}^{(2)}\oplus 2\,\mathbf{3}^{(3)}\oplus 2\,\mathbf{6}^{(0)}$ & $	\mathbf{6}^{(0)}\oplus \mathbf{6}^{(1)}\oplus 2\,\mathbf{3}^{(1)}\oplus 2\,\mathbf{9}^{(0)}$ & $	\mathbf{3}^{(0)}\oplus \mathbf{3}^{(1)}\oplus 2\,\mathbf{15}^{(0)}$ \\ 
        \end{tabular}
}
\end{table*}
The subduction of $ SU(3) $ IRs into those of its finite subgroups are presented in Table~\ref{tab:su3subduction}. Each table row correlates a specific $ SU(3) $ IR, characterized by the quantum numbers $(p, q)$, with a direct sum of the IRs of the finite subgroups. Given that $d_r$ increases with $(p, q)$, satisfying Eq.~\ref{eq:drdj}necessitates larger direct sums.  Further, in contrast to $SU(2)$, where there is only 1 IR of a given dimension, $SU(3)$ can have 0,1 or 2 IRs per dimension.  To compare to the subductions, it is useful to note that $SU(3)$ has one 1d and 8d IR; two 3d, 6d, 10d, 21d IRs; and four 15d IRs.  Thus, looking at the character tables, we can establish that discrete group IRs of other dimensions like $\mathbf{2}$ or $\mathbf{5}^{(0)}$ are subduced only in a direct sum.

From Table~\ref{tab:su3subduction}, we observe that while increasing the group size generically reduces the complexity of the subductions, the number of one-to-one subductions is not always increasing.  Instead, we find that $\sab$ and $\sac$ both have six such subductions.  Further, we observe that for the smallest group $\saa$, the only nontrivial IR that subduces one-to-one is the fundamental.

To further explore the relationship between $SU(3)$ and its subgroups, we again present the inductions. For the sake of conciseness, for pairs of complex IRs we only enumerate the induction for one, e.g. $\mathbf{3}^{(0)}$, while the other in the pair ($\mathbf{3}^{(0)*}$) can be computed by exchanging $p\leftrightarrow q$.  For $\saa$ the induction up to $p+q=5$ is 
\begin{align}
\mathbf{1}^{(0)}&\rightarrow {\{0,0\}|_{1},\{2,2\}|_{1},\{1,4\}|_{1},\{4,1\}|_{1},\hdots}\notag\\
\mathbf{1}^{(1)},\mathbf{1}^{(3)}&\rightarrow {\{0,3\}|_{1},\{3,0\}|_{1},\{1,4\}|_{1},\{4,1\}|_{1},\hdots}\notag\\
\mathbf{1}^{(2)}&\rightarrow {\{2,2\}|_{2},\hdots}\notag\\
\mathbf{3}^{(0)}&\rightarrow \{0,1\}|_{1},\{1,2\}|_{1},\{0,4\}|_{1},\{3,1\}|_{2},\notag \\
&\phantom{xxxxxxxxx}\{2,3\}|_{4},\{5,0\}|_{3},\hdots\notag\\
\mathbf{3}^{(1)},\mathbf{3}^{(3)}&\rightarrow \{2,0\}|_{1},\{1,2\}|_{1},\{0,4\}|_{2},\{3,1\}|_{2},\notag\\&\phantom{xxxxxxxxx}\{2,3\}|_{3},\{5,0\}|_{1},\hdots\notag\\
\mathbf{3}^{(2)}&\rightarrow {\{1,2\}|_{2},\{3,1\}|_{2},\{2,3\}|_{4},\{5,0\}|_{2},\hdots}\notag\\
\mathbf{4}^{(0)},\mathbf{4}^{(0)'}&\rightarrow \{1,1\}|_{1},\{0,3\}|_{1},\{3,0\}|_{1},\{2,2\}|_{3},\notag\\&\phantom{xxxxxxxxx}\{1,4\}|_{4},\{4,1\}|_{4},\hdots\notag
\end{align}
The $\saa$ character table in Tab.~\ref{tab:cts180} shows that every dimension has more IRs than it should to match $SU(3)$ and that the largest pair is 4d.  From the inductions, one observes that the pairs $\mathbf{1}^{(1)}$ \& $\mathbf{1}^{(3)}$, $\mathbf{3}^{(1)}$  \& $\mathbf{3}^{(3)}$, and $\mathbf{4}^{(0)}$ \& $\mathbf{4}^{(0)'}$ always appear together, indication that even when the subduction is not one-to-one that one-to-two subductions can have residual symmetry.  In particular the pair $\mathbf{4}^{(0)}$ \& $\mathbf{4}^{(0)'}$ correspond to the subduction of the adjoint, 8d, IR of $SU(3)$; this IR is commonly used in modified actions for discrete and continuous groups because it arises first in strong-coupling and character expansions~\cite{Bhanot:1981pj,Heller:1995bz,Heller:1995hh,Hasenbusch:2004yq,Hasenbusch:2004wu,Ji:2020kjk,Ji:2022qvr}.  In Sec.~\ref{sec:num}, we investigate the effect of including the 4d IRs into a modified action.  

Another metric to consider is mixing between IRs of $SU(3)$ in the fundamental IR of $\saa$.  Since this IR is typically used in lattice actions, if its mixing under induction represents systematic errors in approximating $SU(3)$ by the group. Assuming dimensionality is a proxy for energy along with the 3d $\{1,0\}$, the next largest contributors to $\mathbf{3}^{(0)*}$ are the 15d $\{2,1\}$ and $\{4,0\}$ IRs.

Moving onto $\sab$, for $p+q\leq5$ the inductions are
\begin{align}
\mathbf{1}^{(0)}&\rightarrow {\{0,0\}|_{1},\{1,4\}|_{1},\{4,1\}|_{1},\hdots}\notag\\
\mathbf{1}^{(1)},\mathbf{1}^{(2)},\mathbf{1}^{(3)}&\rightarrow {\{2,2\}|_{1},\hdots}\notag\\
\mathbf{2}^{(0)}&\rightarrow {\{0,3\}|_{1},\{3,0\}|_{1},\{1,4\}|_{1},\{4,1\}|_{1},\hdots}\notag\\
\mathbf{3}^{(0)}&\rightarrow \{0,1\}|_{1},\{0,4\}|_{1},\{3,1\}|_{1},\notag\\&\phantom{xxxxxxxxx}\{2,3\}|_{2},\{5,0\}|_{2},\hdots\notag\\
\mathbf{3}^{(1)},\mathbf{3}^{(2)},\mathbf{3}^{(3)}&\rightarrow {\{1,2\}|_{1},\{3,1\}|_{1},\{2,3\}|_{2},\{5,0\}|_{1},\hdots}\notag\\
\mathbf{6}^{(0)}&\rightarrow \{2,0\}|_{1},\{1,2\}|_{1},\{0,4\}|_{2},\{3,1\}|_{2},\notag\\
&\phantom{xxxxxxxxx}\{2,3\}|_{3},\{5,0\}|_{1},\hdots\notag\\
\mathbf{8}^{(0)}&\rightarrow \{1,1\}|_{1},\{0,3\}|_{1},\{3,0\}|_{1},\{2,2\}|_{3},\notag\\
&\phantom{xxxxxxxxx}\{1,4\}|_{4},\{4,1\}|_{4},\hdots\notag
\end{align}
Given that $\saa$ is a subgroup of $\sab$, it is interesting to compare their inductions.  First, we observe that the pairs of IRs have merged:  $\mathbf{1}^{(1)}\oplus\mathbf{1}^{(3)}\rightarrow \mathbf{2}$, $\mathbf{3}^{(1)}\oplus\mathbf{3}^{(3)}\rightarrow\mathbf{6}^{(0)}$, and $\mathbf{4}^{(0)}\oplus\mathbf{4}^{(0)'}\rightarrow \mathbf{8}^{(0)}$.  We also see that triplets of $\mathbf{1}^{(1)},\mathbf{1}^{(2)},\mathbf{1}^{(3)}$ and  $\mathbf{3}^{(1)},\mathbf{3}^{(2)},\mathbf{3}^{(3)}$ have arisen. Further, the gaps between $SU(3)$ IRs that mix under induction have grown. For 3d $\{1,0\}\rightarrow \mathbf{3}^{(0)*}$, the nearest IR is now a single 15d one -- $\{4,0\}$.

For $\sac$ which has $\sab$ as a subgroup, the inductions up to $p+q=5$ are
\begin{align}
\mathbf{1}^{(0)}& \rightarrow {\{0,0\}|_{1}\hdots}\notag\\
\mathbf{1}^{(1)}& \rightarrow {\{4,1\}|_{1},\hdots}\notag\\
\mathbf{1}^{(2)}& \rightarrow {\{1,4\}|_{1},\hdots}\notag\\
\mathbf{2}^{(0)}& \rightarrow {\{1,4\}|_{1},\{4,1\}|_{1},\hdots}\notag\\
\mathbf{2}^{(1)}& \rightarrow {\{0,3\}|_{1}\hdots}\notag\\
\mathbf{2}^{(2)}& \rightarrow {\{3,0\}|_{1}\hdots}\notag\\
\mathbf{3}^{(a)}& \rightarrow {\{2,2\}|_{1},\hdots}\notag\\
\mathbf{3}^{(0)}& \rightarrow {\{0,1\}|_{1},\{2,3\}|_{1},\hdots}\notag\\
\mathbf{3}^{(1)}& \rightarrow {\{3,1\}|_{1},\{5,0\}|_{1}\hdots}\notag\\
\mathbf{3}^{(2)}& \rightarrow {\{0,4\}|_{1},\{2,3\}|_{1},\{5,0\}|_{1}\hdots}\notag\\
\mathbf{6}^{(0)}& \rightarrow {\{2,0\}|_{1},\{3,1\}|_{1},\{2,3\}|_{1}\hdots}\notag\\
\mathbf{6}^{(1)}& \rightarrow \{0,4\}|_{1},\{3,1\}|_{1},\{2,3\}|_{1},\{5,0\}|_{1}\hdots\notag\\
\mathbf{6}^{(2)}& \rightarrow {\{1,2\}|_{1},\{0,4\}|_{1},\{2,3\}|_{1}\hdots}\notag\\
\mathbf{8}^{(0)}& \rightarrow {\{1,1\}|_{1},\{2,2\}|_{1},\{1,4\}|_{1},\{4,1\}|_{1},\{3,3\}|_{3},\hdots}\notag\\
\mathbf{8}^{(1)}& \rightarrow \{3,0\}|_{1},\{2,2\}|_{1},\{1,4\}|_{1},\{4,1\}|_{2}\hdots\notag\\
\mathbf{8}^{(2)}& \rightarrow \{0,3\}|_{1},\{2,2\}|_{1},\{1,4\}|_{2},\{4,1\}|_{1}\hdots\notag\\
\mathbf{9}^{(0)}& \rightarrow \{1,2\}|_{1},\{3,1\}|_{1},\{2,3\}|_{2},\{5,0\}|_{1}\hdots\notag
\end{align}
Comparing to $\sab$, we observe that some IRs have merged in $\sac$:  $\mathbf{1}^{(1)}\oplus\mathbf{1}^{(2)}\oplus\mathbf{1}^{(3)}\rightarrow\mathbf{3}^{(a)}$, and  $\mathbf{3}^{(1)}\oplus\mathbf{3}^{(2)}\oplus\mathbf{3}^{(3)}\rightarrow \mathbf{9}^{(0)}$. Unlike the $\saa$ case, some IRs of $\sab$ subduce into new IRs of $\sac$ : $\mathbf{2}^{(0)}\rightarrow\mathbf{2}^{(0)}\oplus\mathbf{2}^{(1)}\oplus \mathbf{2}^{(2)}$, $\mathbf{6}^{(0)}\rightarrow\mathbf{6}^{(0)}\oplus\mathbf{6}^{(1)}\oplus \mathbf{6}^{(2)}$, $\mathbf{8}^{(0)}\rightarrow\mathbf{8}^{(0)}\oplus\mathbf{8}^{(1)}\oplus \mathbf{8}^{(2)}$. The gap between the fundamental IR of  $SU(3)$ and the next has grown, first mixing with the 42-d $\{3,2\}$.

Finally for $\sad$, we have up to $p+q=6$:
\begin{align}
\mathbf{1}^{(0)}& \rightarrow {\{0,0\}|_{1},\{0,6\}|_{1},\{6,0\}|_{1},\hdots}\notag\\
\mathbf{3}^{(0)}& \rightarrow {\{0,1\}|_{1},\{5,0\}|_{1},\{1,5\}|_{1},\hdots}\notag\\
\mathbf{3}^{(1)}& \rightarrow {\{2,3\}|_{1},\{5,0\}|_{1},\hdots}\notag\\
\mathbf{5}^{(0)}&,\mathbf{5}^{(0)'} \rightarrow {\{2,2\}|_{1},\{0,6\}|_{1},\{3,3\}|_{1},\{6,0\}|_{1},\hdots}\notag\\
\mathbf{6}^{(0)}& \rightarrow {\{0,2\}|_{1},\{4,0\}|_{1},\{2,4\}|_{2},\{5,1\}|_{1},\hdots}\notag\\
\mathbf{8}^{(0)}& \rightarrow \{1,1\}|_{1},\{1,4\}|_{1},\{4,1\}|_{1},\notag\\
&\phantom{xxxxxxxxx}\{0,6\}|_{1},\{3,3\}|_{1},\{6,0\}|_{1},\hdots\notag\\
\mathbf{8}^{(0)'}& \rightarrow {\{2,2\}|_{1},\{1,4\}|_{1},\{4,1\}|_{1},\{3,3\}|_{1},\hdots}\notag\\
\mathbf{9}^{(0)}& \rightarrow \{2,2\}|_{1},\{1,4\}|_{1},\{4,1\}|_{1},\notag\\
&\phantom{xxxxxxxxx}\{0,6\}|_{1},\{3,3\}|_{2},\{6,0\}|_{1},\hdots\notag\\
\mathbf{9}^{(1)}& \rightarrow {\{1,3\}|_{1},\{4,0\}|_{1},\{3,2\}|_{1},\{2,4\}|_{2},\{5,1\}|_{1},\hdots}\notag\\
\mathbf{10}^{(0)}& \rightarrow {\{0,3\}|_{1},\{3,0\}|_{1},\{1,4\}|_{1},\{4,1\}|_{1},\{3,3\}|_{2},\hdots}\notag\\
\mathbf{15}^{(0)}& \rightarrow \{1,2\}|_{1},\{3,1\}|_{1},\{2,3\}|_{2},\{5,0\}|_{1},\notag\\
&\phantom{xxxxxxxxx}\{1,5\}|_{2},\{4,2\}|_{2},\hdots\notag
\end{align}
Since $\saa$ is a subgroup of $\sad$, we can compare their inductions, and we observe some of the IRs of $\saa$ have merged in $\sad$:  $5\,\mathbf{1}^{(1)}\oplus 5\,\mathbf{1}^{(3)}\rightarrow\mathbf{10}^{(0)}$,  $\mathbf{4}^{(0)}\oplus\mathbf{4}^{(0)'}\rightarrow \mathbf{8}^{(0)}$, $5\, \mathbf{3}^{(2)}\rightarrow\mathbf{15}^{(0)}$, $\mathbf{3}^{(1)*}\oplus \mathbf{3}^{(2)*}\rightarrow\mathbf{6}^{(0)}$. While generically, the gaps between $SU(3)$ IRs that mix under induction have increased and multiplicities decreased compared to the smaller $\sac$, the fundamental 3d $\{1,0\}$ mixes with the 21-d $\{0,5\}$ rather than the much larger 42-d $\{3,2\}$.  The physical consequences of the difference should be investigated.

\section{Numerical Results}
\label{sec:num}
To study the extent to which discrete subgroups can approximate continuous groups, one can compare expectation values of observables weighted by a lattice action. In general, single-plaquette lattice actions can be constructed from the characters of IRs $\rho$,
\begin{equation}
   S_{\rho}=-\sum_{p,\rho}\beta_{\rho}\Re\chi^{(\rho)}(U_p),
    \label{eqn:WA} 
\end{equation}
where $U_{p}$ is a plaquette built from group-valued links. The commonly-used Wilson action for $SU(3)$ depends only on the fundamental 3d IR: 
\begin{equation}
S_W(\beta)=-\frac{\beta_{\mathbf{3}^{(0)}}}{6}\sum_{p}\Re\chi^{\mathbf{3}^{(0)}}(U_{p}),
    \label{eqn:WA}
\end{equation}
while actions with additional IRs are referred to as \emph{modified actions}. For continuous groups which are asymptotically free, there are typically two "phases" for a lattice action: a lattice phase for which lattice artifacts dominate separated at $\beta_s$ (which in a modified action is a plane of couplings) from a \emph{scaling regime} where these effects are sufficiently small that extrapolation to the continuum can be performed. For finite subgroups, there is an additional Higgs or so-called frozen phase at some subgroup-dependent $\beta_f$ where the discrete subgroup approximation breaks down.  Thus, for a lattice action of a discrete subgroup to be useful for simulations $\beta_s<\beta_f$, i.e., it must support a sufficiently large scaling regime.  From the subductions, we gain a different perspective on how these phases can arise from mixing eigenstates of $SU(3)$ and gain insight into which IRs of the discrete subgroup to add to the action to attempt to improve the agreement with the continuous group. 

Using modified actions has been demonstrated to introduce or enlarge the scaling regime of discrete subgroups~\cite{Bhanot:1981pj,Fukugita:1982kk,Alexandru:2019nsa,Alexandru:2021jpm,Ji:2020kjk,Ji:2022qvr}. The most commonly considered modified action adds the 8d adjoint IR to the Wilson action~\cite{Blum:1996uf,Heller:1995bz,Heller:1995hh,Hasenbusch:2004wu}, and has been demonstrated to reduced the lattice artifacts if the couplings are appropriately chosen.

One useful observable is the expectation value of the plaquette, which corresponds to the average energy density,
\begin{equation}
    E(\beta)=\left\langle 1-\frac{1}{6}\Re\chi^{\mathbf{3}^{(0)}}(U_{p})\right\rangle.
\end{equation}
Comparing to the continuous group $SU(3)$, $E(\beta)$ with the Wilson action is known to agree analytically up to $\mathcal{O}(\beta^4)$ for $\Sigma(1080)$ and $\sac$, while only $\mathcal{O}(\beta^2)$ for the smaller groups $\Sigma(648)$ and $\Sigma(216)$, respectively~\cite{Bhanot:1981xp}. If instead of the Wilson action, an appropriately chosen modified action could be used to improve this strong-coupling expansion -- smoothing left for future work.  Instead, we will consider the nonperturbative calculation of $E(\beta)$ from Monte Carlo simulations.

\begin{figure}
    \centering
    \includegraphics[width=\linewidth]{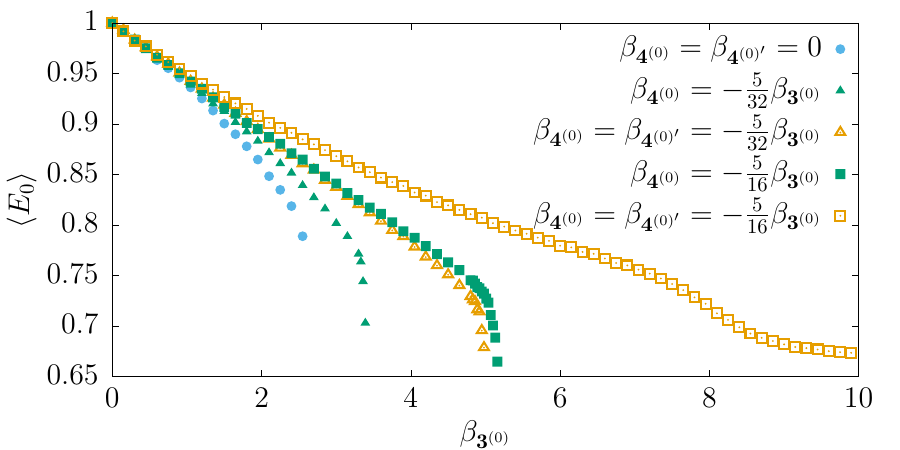}
    \caption{Euclidean calculations of lattice energy density $\langle E_0\rangle$ of $\saa$ as measured by the expectation value of the plaquette for different modified actions as a function of fundamental IR coupling $\beta_{\mathbf{3}^{(0)}}$  on $2^4$ lattices}
    \label{fig:plaqcom}
\end{figure}

Motivated by prior experience that a modified action with the adjoint representation can lead to better agreement with the continuous group, the subgroup $\saa$, which does not have a one-to-one subduction of the adjoint is particularly interesting. In Fig.~\ref{fig:plaqcom}, we plot $E(\beta)$ as a function of $\beta_{\mathbf{3}^{(0)}}$ for different lattice actions. At any given $\beta_{\mathbf{3}^{(0)}}$, the lattice spacing $a$ for different modified actions vary, but the smallness of $E(\beta)$ can be used as a proxy for the discrete subgroup having a larger $\beta_f$.  Further, for $SU(3)$, $E(\beta)\sim 0.6$ corresponds to the scaling regime, so a good discrete subgroup approximation works to get at least that far.  The freezing transition can be recognized by a rapid drop of $E(\beta_f)\rightarrow0$.  In the thermodynamic limit, this will turn into a finite gap at $E(\beta_f)$.

For $\saa$ with the Wilson action, $\beta_{\mathbf{4}^{(0)}}=\beta_{\mathbf{4}^{(0)'}}=0$ we observed the freezing transition occurs when $E(\beta_f)\approx 0.8$.  If we wanted to follow the wisdom of adding something like adjoint representation, we have a choice as to how.  We investigated the relative improvement from adding only $\mathbf{4}^{(0)}$ or $\mathbf{4}^{(0)'}$ or both of them with equal coupling along coupling trajectories: 
\begin{align}
    \beta_{\mathbf{4}^{(0)}}&=-\kappa\beta_{\mathbf{3}^{(0)}},\qquad \beta_{\mathbf{4}^{(0)'}}=0\\
    \beta_{\mathbf{4}^{(0)}}&=\beta_{\mathbf{4}^{(0)'}}=-\kappa\beta_{\mathbf{3}^{(0)}}
\end{align}
for $\kappa=-\frac{5}{32},-\frac{5}{16}$. We see in Fig.~\ref{fig:plaqcom} that for both $\kappa$ including both IRs leads to a smaller value of $E(\beta_f)$.  It is particularly striking for $\kappa=-\frac{5}{16}$ where including both IRs $E(\beta_f)$ never drops to zero. The mechanism for this is related to how $\mathbf{4}^{(0)}$ and $\mathbf{4}^{(0)'}$ give different weights to different $\saa$ group elements.

A deeper investigation of the effects of subduction can be undertaken by looking at the static potentials.  Given that $\sad$ is the best discrete approximation to $SU(3)$, we will focus on this subgroup. The lattice static potential $\hat{V}_\rho$ in a representation is defined with respect to the expectation value of the rectangular Wilson loop in the same representation:
\begin{equation}
\langle W_{\rho}(r,t)\rangle = c_{\rho}(r,a)\text{exp}\left[-\hat{V}_{\rho}(r,a)t\right]\qquad T\rightarrow\infty.    \end{equation}
where $r,t$ are the spatial and temporal extent of the Wilson loop. The prefactor, $c_{\rho}(r,a)$, is an additional fitting parameter that captures aspects of the Wilson loop's behavior at small $r$ that are not related to the linear rise of the potential with $r$; it is extracted when analyzing Wilson loop data to extract the potential.

The static potentials in pure gauge $SU(3)$ theory are expected to following \emph{Casimir scaling} where they are equal for large $r$ up to a rescaling by their quadratic Casimirs~\cite{Bali:2000un,Deldar:1999vi} 
\begin{equation}
C_2(p,q)=\frac{p^2+q^2+3p+3q+pq}{3}.
\end{equation}
Numerical results at both finite lattice spacing and the continuum have confirmed this scaling up to violations smaller than $5\%$.  Here, we compute the static potentials for all unique IRs of $\sad$ and compare their numerical ratios to $R_C=\hat{V}_{\rho}(r/a)/\hat{V}_{\three^{(0)}}(r/a)$ to $C_2(p,q)/C_2(1,0)$ of $SU(3)$.  Given our subduction table, we can check to what extent subducing to a direct sum of IRs differs from subducing to a single IR and how the different IRs of the direct sum behave.

\begin{figure}
    \centering
    \includegraphics[width=\linewidth]{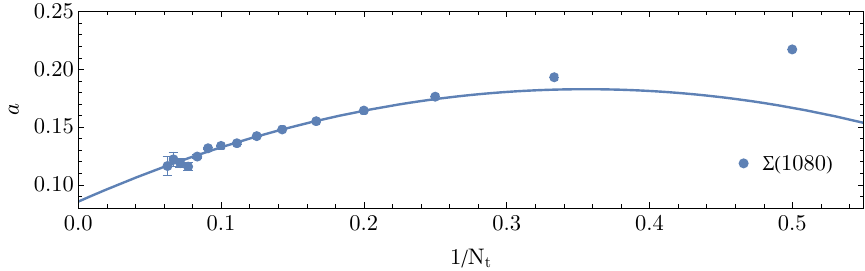}
    \caption{Lattice spacing extracted from the static potential as a function of the temporal extent of the loop $N_t^{-1}$.  The $N_t^{-1}\rightarrow 0$ limit is obtained from a quadratic fit, finding $a=0.085(4)$ fm.}
    \label{fig:latspac}
\end{figure}

\begin{figure*}
    \centering
    \includegraphics[width=\linewidth]{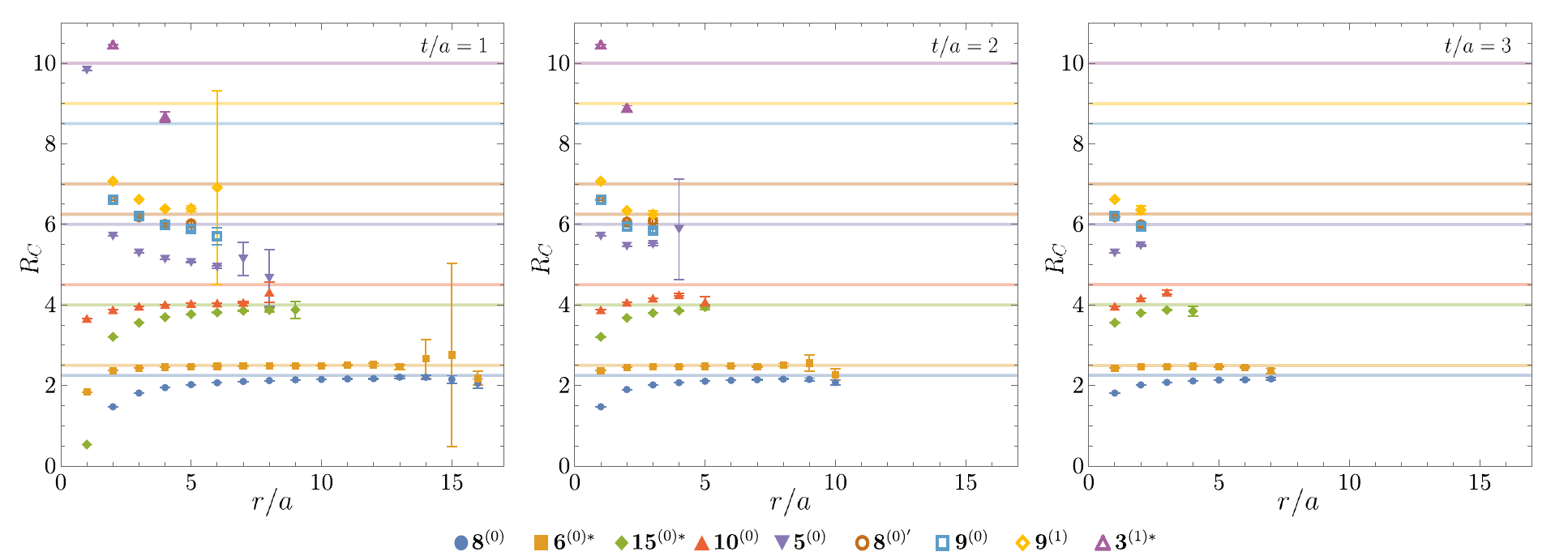}
    \caption{The ratio of static potentials of IRs of $\sad$ to the fundamental $\three^{(0)}$ IR, labelled $R_C$, for all measured representations.  The colored bands correspond to the $SU(3)$ results for all $R_C<10$.}
    \label{fig:statpot}
\end{figure*}

In order to reach the scaling regime, we simulated $\sad$ using a modified action that includes both the fundamental and adjoint IRs
\begin{equation}\label{eq:action-adj}
 S_{fa}=-\sum_p \left(\beta_{\textbf{3}^{(0)}}\Re\chi^{\textbf{3}^{(0)}}(U_p) +\beta_{\textbf{8}^{(0)}}\Re\chi^{\textbf{8}^{(0)}}(U_p)\right)\,,
\end{equation}
With this action, we used $\beta_0=8.407$ and $\beta_1=-1.65$ on a $32^4$ lattice which corresponds to $a=0.085(4)$ fm as determined by using the static potential $\hat{V}_{\three^{(0)}}(r/a)$ with the Sommer parameter $r_0\sim 0.49$ fm defined by~\cite{Sommer:1993ce}
\begin{equation}
    r^2\frac{\partial V}{\partial r}|_{r=r_0}=1.65
\end{equation}
The extrapolation of $a(N_t)$ in the temporal extent of the loop $N_t^{-1}$ to zero is presented in Fig.~\ref{fig:latspac}.
At this lattice spacing, we computed the unique IR static potentials. Because of the nuances involved in adequately accounting for the systematics of smearing discrete subgroups, we have opted not to smear.  As a consequence, the signal-to-noise prevents large $r\times t$ from being probed. Luckily, we observe that similar to prior studies in $SU(3)$~\cite{Bali:2000un,Deldar:1999vi}, $R_C$ appears to plateau rapidly. The results for temporal extents $t=1,2,3$ are presented in Fig.~\ref{fig:statpot} for our ensemble of 32k configurations. 

For the lowest IR, clear signals were obtained for $t/a$ up to 7, and in Tab.~\ref{tab:fits}, we present $R_C(t)$ used to extract infinite limit results.  For IRs where $R_C(t>3)$ were obtained, we extrapolated to the $t\rightarrow\infty$ with a quadratic fit in $t$.  For other IRs where only 2 or 3 values were obtained, for our estimate of $R_C(\infty)$, we take the average and include an additional systematic error, which is the difference between the highest and lowest values.   

\begin{table*}
    \centering
        \caption{Ratios of static potentials of IRs of $\sad$ to the  $\three^{(0)}$ IR, denoted $R_C(\tau)$, where $\tau=t/a$.}
    \begin{tabular}{c|cccccccc}
         $\rho$ & $R_C(1)$ & $R_C(2)$ & $R_C(3)$ & $R_C(4)$ & $R_C(5)$ & $R_C(6)$ & $R_C(7)$&$R_C(\infty)$\\\hline\hline
         $\eight^{(0)}$& 2.2478(10) & 2.2192(14) & 2.2068(28) & 2.205(8) & 2.157(16) & 2.15(7) & 2.01(15)&2.186(14)\\
         $\six^{(0)*}$ & 2.5045(12) & 2.4856(16) & 2.473(4) & 2.478(11) & 2.48(5) & --- & ---&2.463(26)\\
         $\ften^{(0)*}$& 3.978(10) & 4.004(12) & 3.97(9) & --- & --- & --- & ---&3.984(30)(30)\\
         $\ten^{(0)*}$& 4.1522(25) & 4.457(16) & --- & --- & --- & --- & --- &4.305(8)(305)\\
         $\five^{(0)}$& 4.85(5) & 5.84(31) & 5.163(33) & --- & --- & --- & ---&5.28(11)(99)\\
         $\eight^{(0)'}$& 5.82(7)  & 6.44(34) & --- & --- & --- &---  &--- &6.13(17)(62)\\
         $\nine^{(0)}$& 5.26(9) & 5.80(22) & --- &---  & --- & --- & ---&5.53(12)(54)\\
         $\nine^{(1)}$&5.89(17)  &6.3(4)  & --- & --- & --- & --- &--- &6.09(21)(40)\\
         $\three^{(1)*}$&$\lesssim 8.6$  & $\lesssim 9$ & --- & --- & --- & --- &--- &$\lesssim 9$\\\hline
    \end{tabular}
    \label{tab:fits}
\end{table*}

The results for $\sad$ can then be compared to those of the theoretical expectations of $SU(3)=C_2(p,q)/C_2(1,0)$.  Numerical results for $SU(3)$ in ~\cite{Bali:2000un,Deldar:1999vi} have bounded the violation of this scaling to be less than $5\%$, which we take a metric for agreement from the discrete group approximation.  The comparison is shown in \tab{casimir}, where we include the fractional deviation $\delta R$ between the expected scaling and $\sad$.  We observe that for the four IRs of $\sad$, which are subduced one-to-one from $SU(3)$ is $\delta R<5\%$ with two consistent with zero.  In the $\{2,2\}$ case, which subduces into four new IRs, we see mild tension with the $5\%$ violations for the $\mathbb{5}$ IRs, while the $\mathbb{8}^{(0)'}$ and $\mathbb{9}^{(0)}$ are found to be in agreement.  At still higher IRs, we see $\{3,1\}$ and $\{3,2\}$ subduce to direct sums of IRs that include ones for the first time, e.g. $\mathbb{9}^{(1)*}$ and $\mathbb{3}^{(1)*}$, alongside ones that are subduced into lower IRs.  In both cases, we observe that the new IRs are consistent with the anticipated scaling of the $SU(3)$ IR.  Together, these results are suggestive of how the eigenstates of $SU(3)$ break into the IRs of $\sad$ -- and that for the lowest IRs, good agreement between observables should be anticipated.

\begin{table}
    \centering
        \caption{Comparison of the Casimir scaling of the static potentials of $SU(3)$ with to their subducted IRs in $\sad$. In the last column, we present the fraction deviation $\delta R$.}
    \label{tab:casimir}
    \begin{tabular}{clccc}
        $SU(3)$ & $\sad$ & $R_{C}^{SU(3)}$ & $R_{C}^{\sad}$ & $\delta R$ \\
        \hline\hline
        \{1,1\}& $\textbf{8}^{(0)}$ & 2.25 & $2.186(14)$& 0.028(6)\\\hline
        \{2,0\}& $\textbf{6}^{(0)*}$ & 2.5 & $2.46(3)$& 0.016(12)\\\hline
        \{2,1\}& $\textbf{15}^{(0)*}$ & 4 & $3.98(3)(3)$& 0.005(8)(8)\\\hline
        \{3,0\}& $\textbf{10}^{(0)}$ & 4.5  & $4.305(8)(305)$& 0.0433(18)(700)\\\hline
        \{2,2\}& $\textbf{5}^{(0)}$ & 6 &$5.28(11)(99)$& 0.120(18)(70)\\
              & $\textbf{5}^{(0)'}$ &  &$5.28(11)(99)$ & 0.120(18)(70)\\
              & $\textbf{8}^{(0)'}$ &  & $6.13(17)(62)$& 0.022(28)(100)\\
              & $\textbf{9}^{(0)}$ &  & $5.53(12)(54)$& 0.078(20)(90)\\\hline              
         \{3,1\}& $\textbf{9}^{(1)*}$ & 6.25    &$6.09(21)(40)$& 0.026(34)(60) \\
         & $\textbf{15}^{(0)}$ & & $3.98(3)(3)$\footnote{\label{rep} IR present in lower subduction}& 0.363(5)(5)\\\hline      
        \{4,0\}& $\textbf{6}^{(0)}$ & 7 & $2.46(3)$\footref{rep}& 0.649(4) \\
              & $\textbf{9}^{(1)}$ &  &$6.09(21)(40)$\footref{rep}& 0.130(30)(60)\\\hline
        \{3,2\}& $\mathbf{3}^{(1)*}$ & 8.5 & $\lesssim 9$& $\lesssim 0.06$\\
              & $\mathbf{9}^{(1)}$ &   & $6.09(21)(40)$\footref{rep}& 0.284(25)(50)\\
              & $\mathbf{15}^{(0)*}$ & &$3.98(3)(3)$\footref{rep} & 0.5318(35)(35)\\
    \end{tabular}
\end{table}
\section{Conclusion}
\label{sec:con}

We provided the subduction of $SU(N)$ groups of interest in lattice gauge theory to their crystal-like subgroups up to a high dimension. The subduction tables provide new insight into systematically understanding the limitations of the discrete subgroup approximation and give guidance on how to improve the approximation. It was demonstrated how the subduction patterns can be used to determine effective modified actions and how violations of Casimir scaling are signals for the discrete subgroup breakdown. 

For future work, examining how the subduction of continuous groups to their discrete subgroups interacts with the Higgs mechanism could provide valuable insights into symmetry-breaking processes in lattice gauge theories. Further, developing a systematic understanding of the errors in gauge-smearing techniques optimized for discrete subgroups could help reduce signal-to-noise to probe the scaling breakdown more precisely. Additionally, a dedicated analysis of the excited state glueball spectrum within these approximations could provide deeper insights into the stability and spectrum of gauge theories.

\begin{acknowledgements}
HL is supported by the Department of Energy through the Fermilab QuantiSED program in the area of ``Intersections of QIS and Theoretical Particle Physics". This work was supported by the Fermi National Accelerator Laboratory (Fermilab),
a U.S. Department of Energy, Office of Science, HEP User Facility. Fermilab is managed by Fermi Research Alliance,
LLC (FRA), acting under Contract No. DE–AC02–07CH11359.
\end{acknowledgements}

\bibliography{ref}

\end{document}